\documentclass[letterpaper,11pt]{article}

\usepackage[utf8]{inputenc}
\usepackage[T1]{fontenc}

\usepackage{jheppub,multirow, amsthm,mathrsfs,xspace,xcolor,float,color,siunitx,comment}
\usepackage{bm}
\usepackage[caption=false]{subfig}
\usepackage[countmax]{subfloat}
\usepackage{graphbox}

\newtheorem{theorem}{Theorem}
\providecommand{\href}[2]{#2}

\definecolor{darkred}{rgb}{0.5,0.0,0.0}
\definecolor{darkblue}{rgb}{0.0,0.0,0.9}
\definecolor{darkerblue}{rgb}{0.0,0.0,0.5}
\definecolor{darkgreen}{rgb}{0.0,0.5,0.0}
\definecolor{black}{rgb}{0.0,0.0,0.0}
\definecolor{brown}{rgb}{0.6,0.4,0.2}

\DeclareRobustCommand{\Sec}[1]{Sec.~\ref{#1}}

\DeclareRobustCommand{\Fig}[1]{Fig.~\ref{#1}}

\DeclareRobustCommand{\Eq}[1]{Eq.~(\ref{#1})}

\DeclareRobustCommand{\Refs}[1]{Refs.~\cite{#1}}

\title{Systematic Quark/Gluon Identification\\with Ratios of Likelihoods}

\author[a]{Samuel Bright-Thonney,}
\author[c]{Ian Moult,}
\author[d]{Benjamin Nachman,}
\author[e]{and Stefan Prestel}

\affiliation[a]{Physics Department, Cornell University, 109 Clark Hall, Ithaca, New York 14853, USA}
\affiliation[c]{Department of Physics, Yale University, New Haven, CT 06511}
\affiliation[d]{Physics Division, Lawrence Berkeley National Laboratory, Berkeley, CA 94720, USA}
\affiliation[e]{Department of Astronomy and Theoretical Physics, Lund University, S-223 62 Lund, Sweden}

\emailAdd{skb93@cornell.edu}
\emailAdd{ian.moult@yale.edu}
\emailAdd{bpnachman@lbl.gov}
\emailAdd{stefan.prestel@thep.lu.se}

\abstract{
Discriminating between quark- and gluon-initiated jets has long been a central focus of jet substructure, leading to the introduction of numerous observables and calculations to high perturbative accuracy.
At the same time, there have been many attempts to fully exploit the jet radiation pattern using tools from statistics and machine learning.
We propose a new approach that combines a deep analytic understanding of jet substructure with the optimality promised by machine learning and statistics.
After specifying an approximation to the full emission phase space, we show how to construct the optimal observable for a given classification task.
This procedure is demonstrated for the case of quark and gluons jets, where we show how to systematically capture sub-eikonal corrections in the splitting functions, and prove that linear combinations of weighted multiplicity is the optimal observable.
In addition to providing a new and powerful framework for systematically improving jet substructure observables, we demonstrate the performance of several quark versus gluon jet tagging observables in parton-level Monte Carlo simulations, and find that they perform at or near the level of a deep neural network classifier.
Combined with the rapid recent progress in the development of higher order parton showers, we believe that our approach provides a basis for systematically exploiting subleading effects in jet substructure analyses at the Large Hadron Collider (LHC) and beyond.
}

\begin{document} 
\maketitle
\flushbottom

%%%%%%%%%%%%%%%%%%%
\section{Introduction}
\label{sec:intro}
%%%%%%%%%%%%%%%%%%%

Jets are collimated sprays of particles produced by high energy quarks and gluons.  The radiation pattern within jets -- jet substructure -- provides a powerful set of tools for identifying the origin of jets~\cite{Abdesselam:2010pt, Altheimer:2012mn, Altheimer:2013yza, Adams:2015hiv, Larkoski:2017jix,Kogler:2018hem,Marzani:2019hun}.  For example, jet substructure has been used to differentiate between jets originating from quarks and those originating from gluons, as well as to distinguish jets due to highly Lorentz-boosted massive particles (e.g. $W$/$Z$/Higgs bosons or top quarks) and generic quark and gluon-initiated jets.

In most cases, jet classification proceeds by constructing a small set of physically motivated observables.  These observables are often required to have certain properties like infrared and collinear (IRC) safety (or looser requirements like Sudakov safety~\cite{Larkoski:2013paa,Larkoski:2015lea}) so that their cross section can be calculated in perturbation theory.  Furthermore, these observables are usually built to isolate different regions of phase space that are predominantly occupied by one class of jet or another.  In some cases, the decision boundaries are defined using heuristic arguments, while in other applications power counting \cite{Larkoski:2014gra,Larkoski:2014zma,Larkoski:2015kga,Moult:2016cvt,Larkoski:2017iuy,Larkoski:2017cqq,Cal:2022fnm} or other theoretical tools are employed.

While this program has resulted in many interesting physics results, it is not systematically improvable. In particular, a stream of new observables have been proposed over the last decade with increasingly superior numerical performance on particular physics problems.  There is no guarantee that the latest observable is optimal and so there is no natural way of knowing when further observable development is no longer necessary. In parallel to the improvement in observables, there has been significant improvement in the perturbative accuracy of parton showers \cite{Gellersen:2021eci,Dulat:2018vuy,Hoche:2017hno,Li:2016yez,Hoche:2017iem,vanBeekveld:2022zhl,Hamilton:2021dyz,Karlberg:2021kwr,Hamilton:2020rcu,Dasgupta:2020fwr}, incorporating increasingly subtle features into the description of jets that can be exploited for classification tasks. Combined, this suggests that a more systematic approach to observable construction is required. 

In this paper we propose such a systematic approach, that is complementary and opposite to the philosophy commonly taken in the literature.  Instead of first positing an observable and then calculating it to high precision, we propose to specify a given precision and then compute the optimal observable.  Given a probability density\footnote{In practice, these are computed at a given order in perturbation theory and could be negative.  As long as these regions are small and isolated, zeroing them is likely sufficient.} for jets originating from particles $i$ and $j$, the optimal observable for differentiating these types of jets is the likelihood ratio~\cite{neyman1933ix}.  We specifically focus on the universal part of the cross section and consider higher order approximations to the emission phase space density.  While we focus on quark versus gluon jet tagging, the core idea is applicable to a variety of jet tagging tasks. Using this approach, we will show that we can systematically incorporate sub-eikonal corrections to the splitting functions into the observable definition.  The resulting observable itself can be computed to higher orders in perturbation theory, but the observable may no longer be optimal in the sense defined above.

Although our approach yields calculable and systematically improvable approximations of the likelihood ratio, its promise of optimality is strictly limited to the perturbative regime. Any nonperturbative effects -- most notably hadronization in the case of jet physics -- are not accounted for in our calculations, and we will show that they significantly constrain the classification power of our observables. Devising a set of classifying observables that are simultaneously interpretable, (approximately) optimal, and robust to nonperturbative physics is a significant challenge, and falls beyond the scope of this work.

While we find it unlikely that analytic classifiers can completely replace modern machine learning (ML) techniques, particularly those that exploit information beyond energy flow, we still believe that further developing an understanding of the nature of physical information in jets and how it can be exploited for classification tasks is important. First, it can help to reduce the ``black box'' nature of ML techniques, enabling one to gain confidence about the validity of ML techniques in jet physics. Second, it allows one to identify the physical origin of information exploited by the classification task, so that one can ensure that the relevant effects are well modelled, or to motivate further work to improve their description. 

Our approach is related to a number of other proposals in the literature. Most prominently, likelihood-based methods built on cross section calculations underly the Matrix Element Method (MEM) first proposed for top quark mass measurements at the Tevatron~\cite{D0:2004rvt}. However,  the MEM focuses on the hard-scattering part of the cross section and is not readily extendable to other final states.  In the context of jet substructure, an observable similar to the one proposed in this paper is shower deconstruction (SD)~\cite{Soper:2011cr,Soper:2012pb,FerreiradeLima:2016gcz}. Like our optimal observables, SD is built as a likelihood ratio from approximations to the parton shower, however, a key difference is that we are able to translate our result into (an expansion in) standard jet observables.  Other recent work in this direction can be found in \cite{Larkoski:2019nwj,Kasieczka:2020nyd,Bieringer:2020tnw,Dreyer:2021hhr,Lai:2020byl,Buckley:2020kdp,Dreyer:2018nbf}.

This paper is organized as follows.
In \Sec{sec:structure} we provide a review of the structure of quark and gluon jets, emphasizing the organization of the $1\to 2$ splitting functions into eikonal and non-eikonal structures.
In \Sec{sec:squirrel} we describes our approach to constructing optimal classifiers using likelihood ratios, and consider the explicit example of quark versus gluon jet tagging at leading logarithmic (LL) and  modified leading logarithmic (MLL) order.
In \Sec{sec:squirrelemperical} we validate our approach using parton shower (PS) Monte Carlo (MC) simulations, and study the resulting observables.
We conclude in \Sec{sec:conc}.

%%%%%%%%%%%%%%%%%%%%%%%%%%%%%%%%%
\section{The Structure of Quark and Gluon Jets}
\label{sec:structure}
%%%%%%%%%%%%%%%%%%%%%%%%%%%%%%%%%

We begin by analyzing the perturbative structure of a jet sourced by a hard parton radiating massless gauge bosons (the radiation of fermion pairs will be addressed later).
Here, we will restrict ourselves to LL or MLL, where we can use the $1\to 2$ splitting functions. Extensions to higher logarithmic accuracy could be performed by analyzing the $1\to n$ splitting functions. We believe that this is particularly interesting in light of progress in the description of quark and gluon jets with parton shower programs.

While the structure of quark and gluon jets has been extensively discussed in the jet substructure literature, this has almost entirely been from the perspective of the soft-collinear limit, where the only distinction between quarks and gluons is their color charge. However, the $1\to 2$ splitting functions, which are the basis of standard parton shower programs, contain more information. Here we would like to clearly understand the physical nature of the additional information, how it can be exploited, and why it is small compared to the $C_F$ vs. $C_A$ color information.

Although we will ultimately be focused on the physically realized case of QCD, it is interesting to view the problem more generally. This perspective has also been inspired by arguments for the simplification of higher order splitting functions in \Refs{Dokshitzer:2005bf,Dokshitzer:2006nm,Beccaria:2007bb}. We therefore consider the splitting function for the radiation of a massless gauge boson
\begin{align}\label{eq:split_general}
P_i(x)=4C_i \frac{\alpha}{4\pi} \left(\frac{x}{1-x}+(1-x) g_i(x)   \right)\,, \qquad \left\{ \begin{array}{l}g_\phi(x)=0 \\ g_\lambda(x)=\frac{1}{2} \\g_V(x)=x+\frac{1}{x} \end{array} \right. \,.
\end{align}
The function $g_i(x)$ satisfies the condition that it has a regular Taylor series expansion about $x=1$.
Here we see that the splitting probability depends on the parton species through two factors: the Casimir $C_i$ which describes the color charge of the particle, and the function $g_i(x)$.

In the $x\to 1$ limit, in which the emitted radiation is soft, and which is enhanced by the soft singularity, the splitting functions exhibit a universal form depending only on the color factors $C_i$. Explicitly, these are
\begin{align}
C_F=\frac{N^2-1}{2N}\to \frac{4}{3}\,, \qquad C_A=N\to 3\,.
\end{align}
The result in this limit is simply the classical result for radiation in the eikonal limit\footnote{This corresponds with the well-known fact that the leading double logarithmic Sudakov is classical.}~\cite{Levy:1969cr}, which is (as it must be) independent of the spin of the emitting parton.
In this limit, the information in a jet is encoded solely in the \textit{quantity} (multiplicity) of radiation rather than the \textit{structure} of the radiation (we will prove this rigorously in \Sec{sec:squirrel}). Standard studies of quark vs. gluon discrimination have focused solely on this leading eikonal term, resulting in the standard claim that multiplicity or counting observables are optimal (see e.g. Ref.~\cite{Frye:2017yrw}). 

One of the motivations for exploring in more detail the structure beyond the strict eikonal limit, is that there has recently been work on purely collinear jet substructure observables (the projected energy correlators, see \cite{Dixon:2019uzg,Chen:2019bpb,Chen:2020vvp,Chen:2020adz,Chen:2021gdk,Komiske:2022enw,Holguin:2022epo,Chen:2022jhb,Chen:2022swd,Lee:2022ige} for more detailed discussions) that are sensitive to specific integer moments (the twist-2 spin-J anomalous dimensions) of the splitting functions. These observables are not correctly described by the leading eikonal result for the splitting function, and exhibit differences between quarks and gluons generated from the non-eikonal terms in the splitting function. One would therefore like to understand how to exploit this information for quark gluon discrimination.

The non-eikonal term, $g_i(x)$, depends on the nature (spin) of the emitting parton.
Its contribution to the splitting function is suppressed by two powers of $(1-x)$, which is guaranteed by the Low-Burnett-Kroll (LBK) theorem~\cite{Low:1958sn,Burnett:1967km,DelDuca:1990gz} describing the soft limits of gauge theories.\footnote{In this case, the LBK theorem is applied after taking the collinear limit.}
As expected, $g_i(x)$ vanishes for a scalar, which has no structure beyond its color.

Expanding the function $g_i(x)$ in powers of $(1-x)$, we have
\begin{align}
g_\phi=0\,, \qquad g_\lambda=\frac{1}{2}\,, \qquad g_V=\frac{1}{2}+\frac{1}{4}(1-x)^2+\cdots\,.
\end{align}
This shows that in QCD, the difference between a quark and a gluon is in fact more suppressed than it needs to be! This makes the difficulty of quark/gluon jet discrimination abundantly clear: the structure of a jet is dominated by its classical result, with the spin information being highly suppressed. Nevertheless, this information is there, and we should be able to design observables to systematically exploit it.

In QCD, there is an additional distinction between quarks and gluons, namely that gluons can split into a $q\bar q$ pair,
\begin{align}
P_{g\to q\bar q}(z) & = \frac{n_f T_R}{2C_A}[z^2 + (1-z)^2].
\end{align}
This splitting does not obey the LBK theorem and is down only by a single power of $z$.\footnote{See \Refs{Moult:2016fqy,Boughezal:2016zws,Moult:2017jsg,Boughezal:2018mvf,Moult:2018jjd} for a discussion.}

%%%%%%%%%%%%%%%%%%%%%%%%%%%%%%%%%
\section{Quark and Gluon Jet Likelihoods}
\label{sec:squirrel}
%%%%%%%%%%%%%%%%%%%%%%%%%%%%%%%%%

In this section, we introduce an approach which allows us to prove the optimality of quark/gluon discriminants.
In this approach, we analyze the likelihood function of jet radiation at a given accuracy, and apply the Neyman-Pearson Lemma~\cite{neyman1933ix} to construct an optimal discriminant as the likelihood ratio of the two jet classes in question (in our case, quark and gluon jets). Explicitly constructing likelihood ratios has garnered increasing attention in recent years, as it provides some insight into what physical information a deep neural network might be using when learning a (presumed optimal) classification function. Our work is similar in spirit to the approaches in Refs.~\cite{Kasieczka:2020nyd,Dreyer:2021hhr}, and provides a complementary perspective for calculating jet likelihoods.
\par Our technique allows us to work in some approximation (e.g. eikonal, independent emission, etc.) and then analyze the likelihood using machine learning or information-theoretic techniques to prove results about the optimal observables.
These optimality results will then apply to any observables that can be computed at a given accuracy from this likelihood.
Since we will always work in some approximation for the likelihood, one can always come up with observables that can evade the proven results, if they cannot be correctly computed from this approximation.
However, we view this as a virtue, as it shows how results can be systematically improved with more accurate approximations.

%%%%%%%%%%%%%%%%%%%%%%%%%%%%%%%%%
\subsection{Analysis in the Eikonal Limit}\label{sec:eik}
%%%%%%%%%%%%%%%%%%%%%%%%%%%%%%%%%

We begin by considering the optimal quark/gluon discriminant in the eikonal limit. We define the eikonal limit as independent emissions with the universal part of the splitting functions
\begin{equation}
\label{eq:LLemit}
dP_{i\to i g}(z,\theta) = \frac{2 \alpha_s C_i}{\pi} \frac{dz}{z}\frac{d\theta}{\theta},
\end{equation}
where $i\in\{q/g\}$.

Before proceeding, we would like to clarify a number of issues related to our referring to this as the eikonal, or classical eikonal limit.
Another name which could be used is the LL limit.
We avoid this language, since it is best defined when there is an observable of interest.
While it is of course true that the above will generate the leading logarithms for Sudakov-type observables, it is straightforward to identify IRC safe observables (even those that probe only two particle correlations) for which the leading logarithm will not be produced by the above approximated splitting function. 

Since we have shown in \Sec{sec:structure} that the splitting function in this limit is identical for all partons up to the color factor, we should be able to prove that multiplicity is the optimal observable.
To show this, we consider the quark/gluon likelihood ratio $L_{q/g}$ for radiating a collection of gluons\footnote{Here we use the well known fact that there is a probabilistic interpretation for the twist two splitting functions. This, however, should not be taken for granted, and indeed such a probabilistic interpretation fails at higher twist \cite{Jaffe:1982pm}.} with kinematics $\{(z_n,\theta_n)\}_{n=1}^M$ 
\begin{equation}\label{eq:LqgLL}
\ln L_{q/g}^\text{LL} = \sum_{n=1}^M \ln \left(\frac{dP_{q\to qg}(z_n,\theta_n)}{dP_{g\to gg}(z_n,\theta_n)} \right)= \ln \frac{C_F}{C_A} \sum_{n=1}^M 1 = M \ln \frac{C_F}{C_A}.
\end{equation}
Here $M$ is simply the multiplicity, and so by the Neyman-Pearson Lemma\footnote{Monotonic functions of the likelihood ratio are also optimal since monotone functions do not change the ordering by the classifier.}~\cite{neyman1933ix}, the optimal observable at this level of accuracy is the multiplicity.
Experimentally the multiplicity is indeed found to perform well as a discriminant~\cite{CMS:2013kfa,ATLAS:2016wzt}, and we provide justification for why this is the case.

Another perspective is that the optimality of counting follows simply from the universal nature of soft emissions in gauge theory, or in other words, the classical structure of the jet.
This behavior is true for a very wide class of observables, including standard Sudakov observables, Sudakov observables that have been soft dropped \cite{Larkoski:2014wba}, and groomed multiplicity \cite{Frye:2017yrw}.
We also note that the above result holds regardless of whether the coupling is taken to be running or not, since it cancels out of the ratio in \Eq{eq:LqgLL}. We therefore state our first result:
\begin{theorem} 
For an observable whose LL result can be computed using the eikonal splitting functions of \Eq{eq:LLemit} in the independent emission approximation (with or without running coupling), this LL result can not achieve better quark/gluon discrimination than multiplicity.
\end{theorem}

\begin{figure}[t]
\centering
\includegraphics[width=0.5\textwidth]{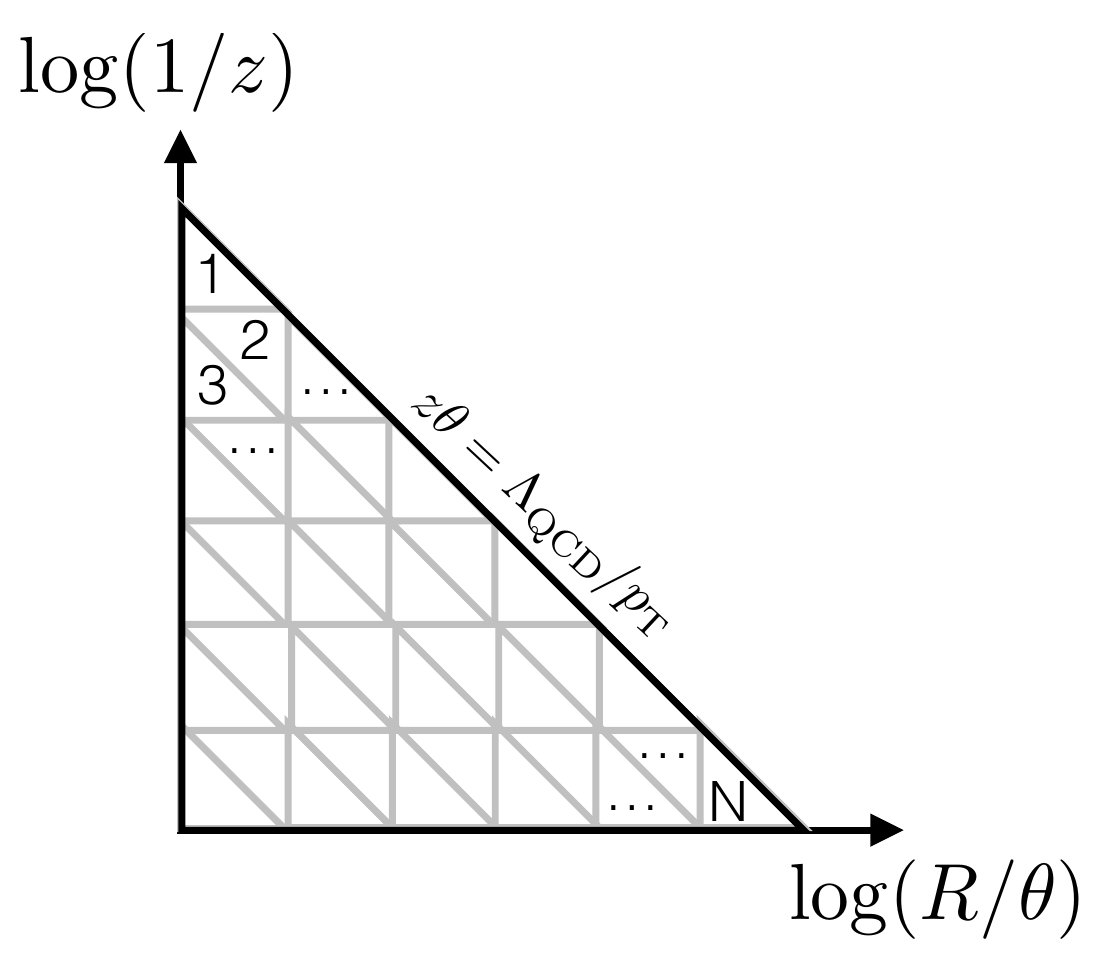}
\caption{A tiled version of the Lund plane where each of the $N$ triangles has the same area.  The non-perturbative regime is defined by $z\theta>\Lambda_\text{QCD}/p_\text{T}$.}
\label{fig:lundschematic}
\end{figure}

In addition to showing that multiplicity is optimal, we can in fact derive its probability distribution using Poisson distributions for the multiplicity $M$\footnote{Poisson observables were emphasized in \cite{Frye:2017yrw}.}
\begin{equation}
p^\text{LL}_i(M) = \text{Pois}\left[\int \frac{dz}{z} \int \frac{d\theta}{\theta}\, \frac{2\alpha_s C_i}{\pi} \Theta(z,\theta)\right]\,.
\end{equation}
In the eikonal limit, emissions are uniformly distributed in the Lund plane, and we can tessellate some perturbative region of the plane using triangles (see \Fig{fig:lundschematic}).
The probability for $n$ emissions in a given triangle with area $\Delta$ is then Poisson distributed according to
\begin{align}
\Pr(n_i=n)=\frac{\lambda^n e^{-\lambda}}{n!}\,, \qquad \lambda=\frac{2\alpha_s C_i \Delta}{\pi}\,.
\end{align}
Due to the assumption of independent emissions, we then have that the radiation counts throughout the emission plane are distributed according to
\begin{align}
{\Pr}_i(n_1, n_2, \cdots, n_N)=\prod\limits_{j=1}^N \frac{\lambda_i^{n_j} e^{-\lambda_i}}{n_j!}\,.
\end{align}
The optimal quark/gluon discriminant is then
\begin{align}
\ln \frac{{\Pr}_q(n_1, \cdots, n_N)}{{\Pr}_g(n_1, \cdots, n_N)}  = \ln \frac{C_F}{C_A} \sum\limits_{j=1}^N n_j  + \text{const}.
\end{align}
We can take the $\Delta\to 0$ limit to find again that the optimal quark gluon discriminant is simply the multiplicity.
This version of the proof makes clear that we can apply a cutoff and only consider the likelihood function in the perturbative regime.

%%%%%%%%%%%%%%%%%%%%%%%%%%%%%%%
\subsection{Systematic Expansion Beyond the Eikonal Limit}
%%%%%%%%%%%%%%%%%%%%%%%%%%%%%%%

The radiation phase space from the previous section can be systematically improved.  Observables computed with modified leading logarithmic accuracy should include the effects of the running coupling as well as subleading terms in the splitting function.  In particular, the radiation phase space is given by 
\begin{equation}
\label{eq:LLemit2}
dP_{i\to i j}(z,\theta) = \frac{2\alpha_s (z\theta p_T) \, C_i}{\pi} p_{i\to i j}(z) dz\,\frac{d\theta}{\theta},
\end{equation}
where $p_{ij}$ are the QCD splitting functions.  In general, at this beyond-Eikonal (BE) order, there are can be flavor changes between quarks and gluons, since $\int_{1/2}^1 dz p_{q\to qg}(z) > 0$ and a finite part of the gluon splitting function contains a contribution from $g\rightarrow q\bar{q}$.  Ignoring flavor changing\footnote{See e.g. Ref.~\cite{Medves:2022ccw} for a careful treatment of flavor changing effects.}, the formalism from the previous section still holds, only now the optimal observable is more complicated than simply multiplicity.   In particular:
\begin{equation}
\label{eq:MLLLL}
\text{JL}^\text{BE'} \equiv \ln L_{q/g}^\text{BE'} = \sum_{n=1}^M \ln \frac{dP_{q\to qg}(z_n,\theta_n)}{dP_{g\to gg}(z_n,\theta_n)} = \sum_{n=1}^M\left[\ln \frac{C_F}{C_A} + \ln \frac{p_{q\to qg}(z_n)}{p_{g\to gg}(z_n)} \right],
\end{equation}
where BE' denotes the beyond Eikonal approximation for the radiation phase space, but ignoring flavor changing.  By inserting the corresponding the following splitting functions\footnote{The $p_{q\to qg}$ still includes the flavor change, but this fact is ignored in the BE' approximation.}
\begin{align}
p_{q\to qg}(z) & = \frac{1 + (1 - z)^2}{2z},\\
p_{g\to gg}(z) & = \frac{1 - z}{z} + \frac12 z(1-z),
\end{align}
the optimal observable is
\begin{align}
\label{eq:BEprime}
\text{JL}^\text{BE'} = \sum_{n=1}^M\left[\ln \frac{C_F}{C_A} + \ln \frac{1+(1-z_n)^2}{(1-z_n)(2 + z_n^2)} \right]\approx M\ln \frac{C_F}{C_A}+\frac{1}{2}\sum_{n=1}^M z_n^3+\mathcal{O}(z_n^4).
\end{align}

Equation~\ref{eq:BEprime} shows that the multiplicity is still nearly optimal when including the non-flavor changing corrections to the radiation phase space.  %A full BE calculation requires a proper accounting of flavor changing.  This is accomplished in two steps.  As a first step, we consider a new basis of particles in which there no flavor changes.  For a quark or gluon at a given step in the jet formation, the flavor transition matrix is given by

%\begin{align}
%\begin{pmatrix}q \cr g\end{pmatrix}\mapsto \begin{pmatrix}p(q|q) & p(g|q) \cr p(q|g) & p(g|g)\end{pmatrix}\begin{pmatrix}q \cr g\end{pmatrix},
%\end{align}
%
%\noindent where
%
%\begin{align}
%p(q|q)&=\int_0^{1/2} p_{q\to qg}(z) dz=\frac{1}{2}\int_0^{1/2} dz\left(\frac{1 + (1 - z)^2}{(z)_+}\right)=\frac{1}{2}\int_0^{1/2}dz\frac{(1 - %z)^2}{z}\cr
%p(g|q)&=\int_{1/2}^{1} p_{q\to qg}(z) dz=\int_{1/2}^1 dz \frac{1 + (1 - z)^2}{2z}= \cr
%p(g|g)&=\int_{0}^{1} p_{g\to gg}(z) dz=\int_0^1 dz \frac{1 - z}{z} + \frac12 z(1-z)= \cr
%p(q|g)&=\int_{0}^{1} p_{g\to qq}(z) dz=\int_0^1 \frac{n_f T_R}{2C_A}[z^2 + (1-z)^2] = 
%\end{align}
%
%\clearpage
%
%\begin{equation}
%\label{eq:MLLLL}
%\text{JL}^\text{MLL} \equiv \ln L_{q/g}^\text{MLL} = \sum_{n=1}^M \ln \frac{dP_{q\to qg}(z_n,\theta_n)}{dP_{g\to gg}(z_n,\theta_n)} = %\sum_{n=1}^M\left[\ln \frac{C_F}{C_A} + \ln \frac{P_q(z_n)}{P_g(z_n)} \right].
%\end{equation}
%
%Can see at this point that the optimal observable is a linear combination of the weighted multiplicities considered in 
A full BE calculation requires a proper accounting of flavor changing, which is omitted here. When summed over final states, the quark and gluon splitting functions become
\begin{align}
P_q(z) & = \frac{1 + (1 - z)^2}{2z},\\
P_g(z) & = \frac{1 - z}{z} + \frac12 z(1-z) + \frac{n_f T_R}{2C_A}[z^2 + (1-z)^2].
\end{align}
where the $z\leftrightarrow 1-z$ symmetry is used put the gluon singularity entirely at $z\to 0$. The splitting functions influence the optimal observable via the log likelihood ratio $\ln P_q(z)/P_g(z)$, giving 
\begin{align}
\label{eq:optobsMLL}
\text{JL}^\text{MLL} & = \sum_{n=1}^M\left[\ln \frac{C_F}{C_A} + \ln \frac{1+(1-z_n)^2}{(1-z_n)(2 + z_n^2) + \frac{n_f T_R}{C_A}z_n(z_n^2 + (1-z_n)^2)} \right].
\end{align}
Expanding in powers of $z_n$, we have
\begin{align}
\text{JL}^\text{MLL} &= \sum_{n=1}^{M}\left(\ln \frac{C_F}{C_A} - \frac{n_f T_R}{2C_A} z_n + \frac{n_f T_R (4 C_A + n_F T_R)}{8 C_A^2} z_n^2 + \cdots \right)
\\& = \ln \frac{C_F}{C_A} n^{(\kappa = 0 )} - \frac{n_f T_R}{2C_A} n^{(\kappa = 1)} + \frac{n_f T_R (4 C_A + n_F T_R)}{8 C_A^2} n^{(\kappa = 2)} + \cdots,
\end{align}
where
\begin{equation} \label{nkappa}
    n^{(\kappa)} = \sum_{n=1}^M z_n^\kappa. 
\end{equation}
We can therefore extend to our second result:
\begin{theorem}
For an observable whose LL or MLL result can be computed using splitting functions in the independent emission approximation, this result can not achieve better quark/gluon discrimination than a linear combination of weighted multiplicities.
\end{theorem}

%%%%%%%%%%%%%%%%%%%
\subsection{Perturbative Multiplicity and Parton Showers}
\label{sec:ps_multis}
%%%%%%%%%%%%%%%%%%%

Parton showers are numerical tools that generate multi-parton 
phase-space points on which observables can be measured. The shower proceeds 
iteratively, producing its own initial conditions for subsequent evolution.
This procedure requires physically sensible intermediate states, i.e.\ on-shell 
phase-space points during its ordered evolution. Parton showers recover 
leading-logarithmic results for observables that are sufficiently similar to 
their ordering variable, and for which a strong ordering of emissions is 
guaranteed. The latter aspect and the desire for physically sensible 
intermediate states can sometimes lead to conflicting needs, especially for
observables depending/relying on states containing several partons that cannot
be interpreted as ``ordered", and that are sufficiently different from
the ordering variable. In such observables, kinematic recoil effects can no 
longer be neglected, since different schemes to ensure momentum conservation 
become distinguishable. Perturbative multiplicity is such an observable.

One particularly common scheme to enforce momentum conservation arises from
the leading-color approximation of QCD, in which color-dipoles radiate soft
gluons coherently. This dipole picture suggests to produce physical 
$(n+1)$-parton states from $n$-parton states by distributing the momenta
of two originator partons over three new partons, thus allowing 
momentum-conservation and on-shell conditions throughout. The originator
pair is, in most cases, determined by color connections. Quarks and antiquarks
will thus contribute to the evolution of a single dipole, while gluons 
will participate in two dipoles. 

In parton showers, quarks and gluons yield different radiation patterns (and
thus perturbative multiplicities) because of differing color factors, and
because of different phase-space dependence of hard-collinear contributions.
Moreover, the effect of kinematic recoil is handled differently for both. This
is easily seen by comparing the evolution of $Z\rightarrow q\bar q$ to 
$h\rightarrow g_1\bar g_2$. In the former, the original color connection between $q$ and 
$\bar q$ is broken after the first gluon emission, while for the latter, only
one of the two original connections is severed by emitting the first gluon.
If momentum is re-distributed locally within color-connected dipoles (as
is commonly the case in parton showers, including the \textsc{Dire} shower~\cite{dire} 
employed below), no subsequent gluon emission from the $q$ will influence 
the $\bar q$ momenta, while emissions from $g_1$ can continue to influence the 
$g_2$ momentum, until the second original connection is severed.
This simple choice of recoil mechanism has the advantage that it makes the action
of the parton shower locally invertible, and allows stopping and restarting the
shower at will without consequences. These factors are crucial for matching 
and merging methods to improve the overall event generator fidelity. 
However, such a recoil scheme clearly complicates the subleading-color
behavior of the parton shower, and can lead to unphysical artifacts in 
quark-vs-gluon discrimination. 

To assess the impact of recoil on perturbative multiplicities, we introduce
a simple extension of the local recoil strategy of \textsc{Dire}, with the aim
to obtain an identical recoil handling for quarks and gluons. For this, we 
introduce a ``backbone'' dipole for gluon evolution, which is never allowed
to split. The backbone dipole is defined by the left-over original
color-connection after the first gluon emission in the shower. If the
original state consisted of a $q\bar q$ pair, no backbone is present. The
emission rate from the backbone dipole is forced to vanish. If a parton
contributes both to the backbone and to another dipole evolution (i.e.\ is
a gluon), then the rate of emissions from the parton in the latter dipole is 
increased to ensure that the overall $g\rightarrow gg$ rate is recovered.
This guarantees that recoil effects in $Z\rightarrow q\bar q$ and
$h\rightarrow g_1\bar g_2$ are treated completely identically, by effectively
treating the recoil of emissions from gluons connected to the backbone identical
to emission from (anti)quarks.

%%%%%%%%%%%%%%%%%%%
\subsection{Discussion and Extensions}
\label{sec:discuss}
%%%%%%%%%%%%%%%%%%%
In this section, we introduced a strategy for constructing optimal discriminating observables using quark/gluon jet likelihoods at fixed perturbative accuracy. In the next section, we demonstrate this strategy for quark/gluon jet discrimination in parton-level Monte Carlo events generated at a fixed order. While this technique sheds new light on the quark/gluon jet classification problem in particular, it need not be restricted to this application. It can be applied whenever one can compute class likelihoods for an observable, and performance can be systematically improved to track with advances in theoretical calculations. 

A key advantage of our approach is that it can go systematically beyond the leading eikonal limit that has been the focus of most studies of quark-versus-gluon discrimination. We envision that this could be particularly interesting for understanding how information in higher order splitting functions can be exploited for quark gluon discrimination.

However, we must note that our approach relies on having an understanding of the underlying process describing the formation of quark and gluon jets. While this is true in perturbation theory, it is not true for the hadronization process, for which their exists little analytic understanding. Therefore while we find it exciting that we are able to systematically derive optimal observables from a given set of splitting functions, we must be cautious, particularly for subleading information, that it may be washed out or dominated by hadronization effects. We will study this in \Sec{sec:duffPS}.

%%%%%%%%%%%%%%%%%%%
\section{Monte Carlo Studies}
\label{sec:squirrelemperical}
%%%%%%%%%%%%%%%%%%%
In the previous section, we introduced a systematic method to compute likelihood ratio observables (LROs) corresponding to (approximately) optimal quark/gluon jet discriminants at a given perturbative accuracy. In this section, we will study their classification performance using parton-level Monte Carlo simulations of quark and gluon jets. We alter the functional form of $q\to qg$ emissions in a leading log (LL) shower, producing a simple class of LROs $\ln(P_q/P_g)$, and compare their classification performance to several benchmarks: jet multiplicity and family of deep neural networks (DNNs). When sufficiently trained, DNNs are often presumed to classify optimally, and thus provide an approximate performance ceiling for assessing LROs.

\subsection{Configurable Parton Showers}
\label{dire}
We generate parton-level Monte Carlo events using \textsc{Pythia 8.303} \cite{pythia82} with the \textsc{DIRE} parton shower \cite{dire}. \textsc{DIRE} features a number of settings (``kernel orders'') for the perturbative order of the shower, the simplest of which corresponds to a LL shower without $g \rightarrow q\bar{q}$ splittings (\texttt{KernelOrder = -1}). Accounting for $g \rightarrow q\bar{q}$ splittings significantly complicates calculating the quark/gluon jet likelihoods, so we use \texttt{KernelOrder = -1} for all studies presented here. Pure samples of quark and gluon jets are extracted from $e^+e^- \rightarrow H \rightarrow q\bar{q}/gg$ events, respectively, generated at $\sqrt{s} = m_H = 250$ GeV and with all initial-state radiation turned off. In each event, two sets of jets are clustered using the anti-$k_T$ algorithm with radii $R = 0.4$ and $R = 1.0$ \cite{Cacciari:2008gp}. To minimize the chance of including secondary jets from wide-angle emissions, we only analyze the leading jet from each event.

\par In addition to fixing the shower order, \textsc{DIRE} allows us to tweak the functional form of the splitting functions $P_{q\to qg}$ and $P_{g\to gg}$. In our study, we consider splittings of the form
    \begin{equation}
    \label{eq:diresplits}
	\begin{aligned}
		P_{g \to gg} &\propto C_A \times f_\mathrm{univ}(z)\\ 
		P_{q\to qg} &\propto C_F\times c_0\exp(c_1z + c_2z^2) \times f_\mathrm{univ}(z),
	\end{aligned}
	\end{equation}
	where $c_0$, $c_1$, and $c_2$ are tunable parameters and $f_\mathrm{univ}(z)$ is the universal part of the splitting function. In this case, the log likelihood ratio takes on a particularly simple form:
	\begin{equation}\ln \left(\frac{P_{q\to qg}}{P_{g\to gg}}\right) = \ln\left(c_0\frac{C_F}{C_A}\right) + c_1z + c_2z^2.\end{equation}
	On the level of jets, this defines a likelihood ratio observable
	\begin{align} \label{MCObs}
		\begin{split}
		\ln\mathcal{L} &= \sum_{i\in \mathrm{jet}}\ln\left(c_0\frac{C_F}{C_A}\right) + c_1z_i + c_2z_i^2 \\
		&= \ln\left(c_0\frac{C_F}{C_A}\right)n^{(0)} + c_1n^{(1)} + c_2n^{(2)}\,,
		\end{split}
	\end{align}
	where $n^{(i)}$ is defined as in Eq.\ \ref{nkappa} with $z_i = p_{T,i}/p_{T,\mathrm{jet}}$.
	In our studies, we vary the three parameters above, as well as the color factor $C_F$, and track how the LRO performs as a quark/gluon jet classifier.
	
\subsection{Optimality \& Particle Flow Networks}
\label{sec:pfns}
The family of LROs defined in Sec.\ \ref{dire} are only \textit{approximately} optimal, as the calculation makes the simplifying assumptions that (a) all emissions $z_i$ come from the initiating hard parton, and (b) the momentum fractions $z_{i,\text{jet}}$ relative to the \textit{jet} $p_T$ are sufficient proxies for the fractions $z_i$ relative to the emitting partons. In this study we compare LROs to Particle Flow Networks (PFNs), which use a per-particle latent space embedding to learn arbitary permutation-invariant jet substructure observables \cite{NIPS2017_f22e4747,Komiske:2018cqr,Komiske:2017aww}. We use the constituent kinematic features $(z_i,\eta_i,\phi_i)$ as the per-particle inputs,\footnote{We use pseudo-rapidity $\eta$ rather than rapidity, as it is a standard observable in collider experiments} where momentum fractions are defined as $z_i = p_{T,i}/p_{T,\mathrm{jet}}$ and $(\eta_i,\phi_i)$ are measured relative to the jet axis.\footnote{The jet axis is defined as $(\hat{\eta},\hat{\phi}) = \sum_{i \in \mathrm{jet}}p_{T,i}(\eta_i,\phi_i)/p_{T,\mathrm{jet}}$} The particle embedding map $\Phi: \mathbb{R}^3 \to \mathbb{R}^\ell$ and binary classification function $F: \mathbb{R}^\ell \to \mathbb{R}^2$ are trained simultaneously as fully connected neural networks. Jets are represented in the latent space by summing their constituent particle representations, and the sum is classified by the function $F$.

\par In addition to the PFN we train an Energy Flow Network (EFN), which learns an explicitly infrared and collinear (IRC) safe observable of the form $F(\sum_{i\in \mathrm{jet}}z_i\Phi(\eta_i,\phi_i))$. The EFN provides complementary information to the PFN, indicating whether IRC-safety restricts access to useful classification information (multiplicity, for instance, is IRC unsafe and thus could not be directly learned by an EFN). We also train two variants of a PFN --  $\mathrm{PFN}[z]$ and $\mathrm{PFN}[\eta,\phi]$ -- which use only a subset of the per-particle information, indicated by the variables in brackets. This split elucidates how much useful information comes from the angular structure of the jet versus solely the momenta. The LROs depend only on $z_i$, so comparing their performance against $\mathrm{PFN}[z]$ will indicate if they are optimal given the restricted set of substructure information.

\par We implement all networks with the tools provided in the \texttt{EnergyFlow} package \cite{Komiske:2017aww,Komiske:2018cqr}, based on \texttt{TensorFlow}~\cite{tensorflow}/\texttt{Keras}~\cite{keras}/\texttt{Adam}~\cite{adam}, and use architectures that have been shown to perform well for quark/gluon jet classification \cite{Komiske:2018cqr}. In particular, the particle embedding $\Phi$ is trained as a fully connected network with layer widths of $(100,100,128)$ (where $\ell = 128$ is the latent space dimension). The output mapping has hidden dimensions $(100,100,100)$ and an output dimension $2$, with the Euclidean unit vectors $\hat{e}_1,\hat{e}_2$ acting as quark/gluon jet truth labels. To avoid overfitting, a uniform dropout rate of 20\% is applied to all nodes in $F$. To train each network, a sample of 1 million quark jets and 1 million gluon jets is combined, shuffled, and split into training (70\%), validation (15\%), and testing (15\%) sets. The networks are trained and validated on the corresponding sets over three epochs\footnote{Due to the large number of events ($\sim 1.4$M) in the training set, model training converged very quickly}, and the receiver operating characteristic (ROC) curve and area under the curve (AUC) metrics are evaluated on the test set. We use these metrics to compare network performance with the predicted likelihood ratio observable.

\subsection{Results: \textsc{Pythia} \& \textsc{DIRE}}
\label{sec:results}
The central focus of our studies is varying the parameters $c_1$ and $c_2$ of Eq.\ \ref{MCObs}. We retain the default setting $c_0 = 1$, as any variation of $c_0$ corresponds to a term proportional to multiplicity in the likelihood ratio. We instead vary $C_F$ between its default value $C_F = 4/3$ and $C_F = C_A = 3$. Setting $C_F = 3$ removes the multiplicity term from Eq.\ \ref{MCObs}, leaving it sensitive solely to the higher order terms $n^{(1)}$ and $n^{(2)}$. Using $C_F = 4/3$ reintroduces the $n^{(0)}$ term, making the other two sub-leading effects. Furthermore, we only consider $(c_0,c_1) \in \{0,1\}^2$, since these settings are sufficient to add or remove the corresponding terms in Eq.\ \ref{MCObs}.

\subsubsection{Parton Shower Validation}
\label{sec:CFscan}
\begin{figure}[t]
	\centering
	\includegraphics[width=0.49\textwidth]{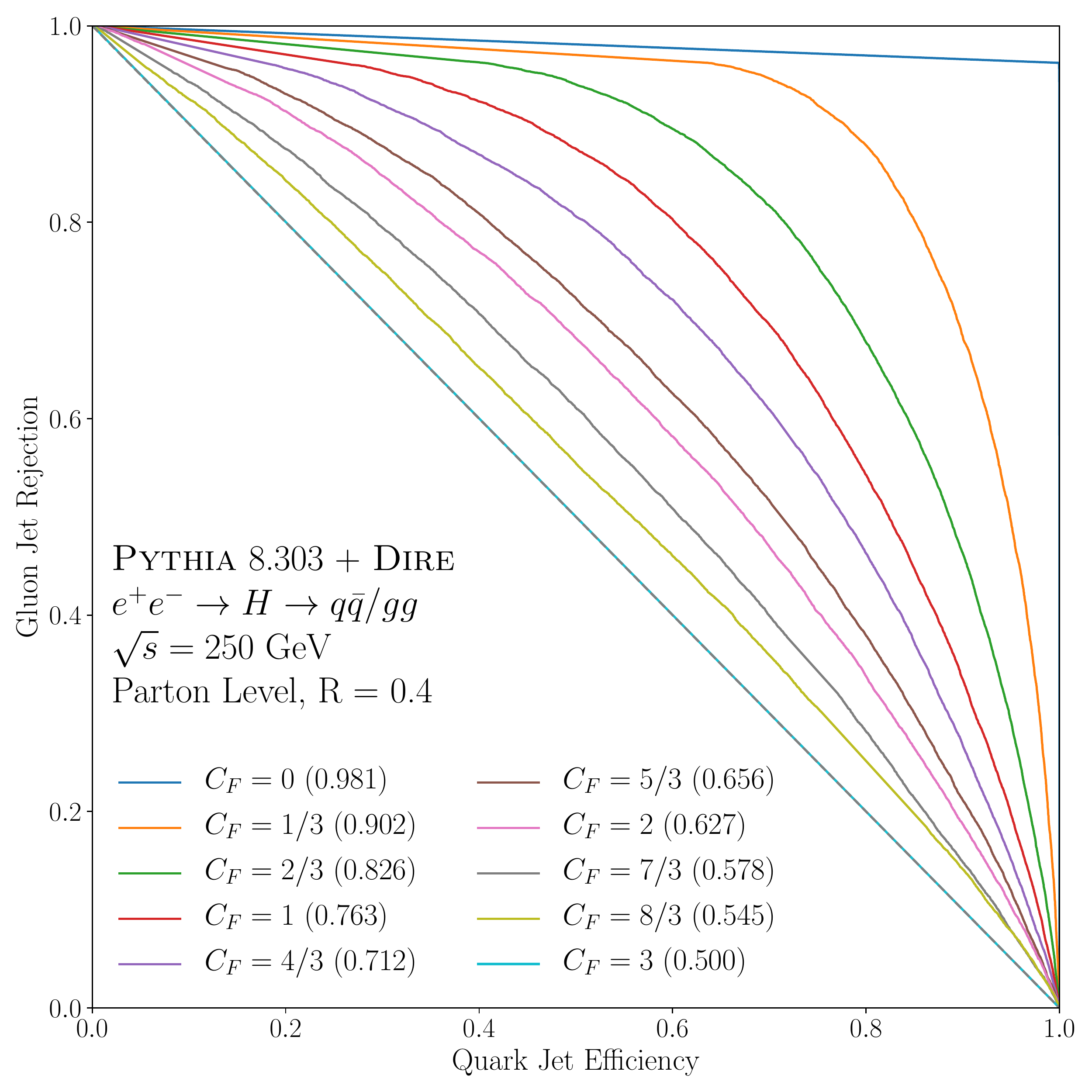}
	\caption{ROC curves for PFN quark/gluon jet classification over a range of quark color factors $C_F$. AUC scores are shown next to each legend entry. At each $C_F$ setting, $q \to qg$ and $g \to gg$ splitting functions are configured to be identical up to the color factor. As expected, quark and gluon jets are indistinguishable at $C_F = C_A = 3$ and nearly perfectly distinguishable at $C_F = 0$.}
	\label{fig:CFscan}
\end{figure}

We begin by validating the expected behavior of the \textsc{DIRE} parton shower and PFN training procedure. Since we are working at \texttt{KernelOrder = -1}, we expect quark jets and gluon jets to look identical when $C_F = C_A$. In Fig.\ \ref{fig:CFscan}, we show ROC curves and AUC scores for PFN quark/gluon jet classifiers trained on quark jets generated with a range of $C_F$ values between $C_F = 0$ to $C_F = 3$\footnote{The same set of gluon jets generated with $C_A = 3$ is used in each training.}. As expected, the two are indistinguishable when $C_F = 3$. This confirms two crucial details for our later studies: first, that the $q \to qg$ and $g \to gg$ splitting functions can be made identical by tuning $C_F$, and second, that our PFN architecture and training protocol is not overfitting the limited amount of information available in the parton-level shower.

\subsubsection{Exponential Modifications of $q \to qg$}
\label{sec:expscan}
\begin{figure}[t]
	\centering
	\includegraphics[width=0.49\textwidth]{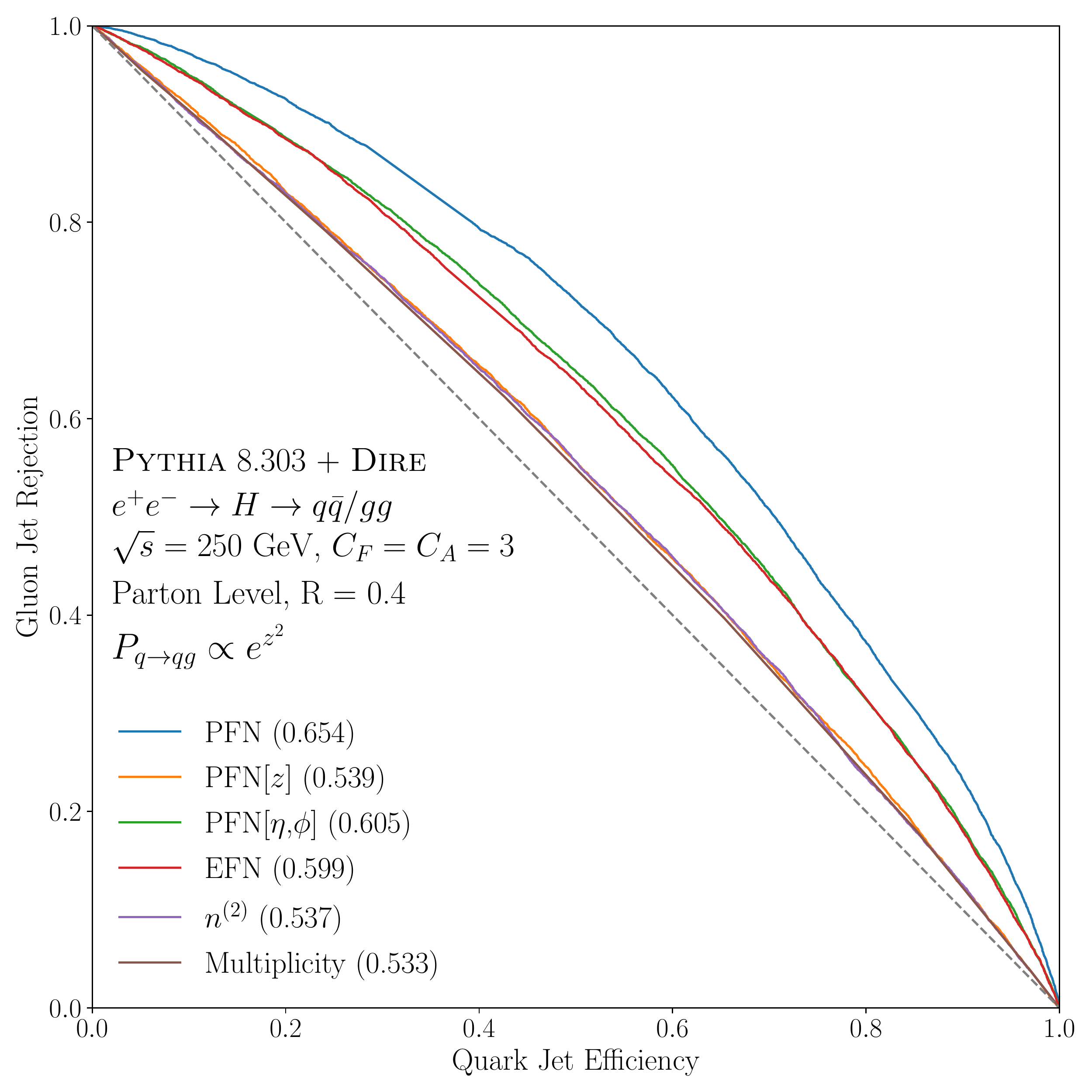}
	\includegraphics[width=0.49\textwidth]{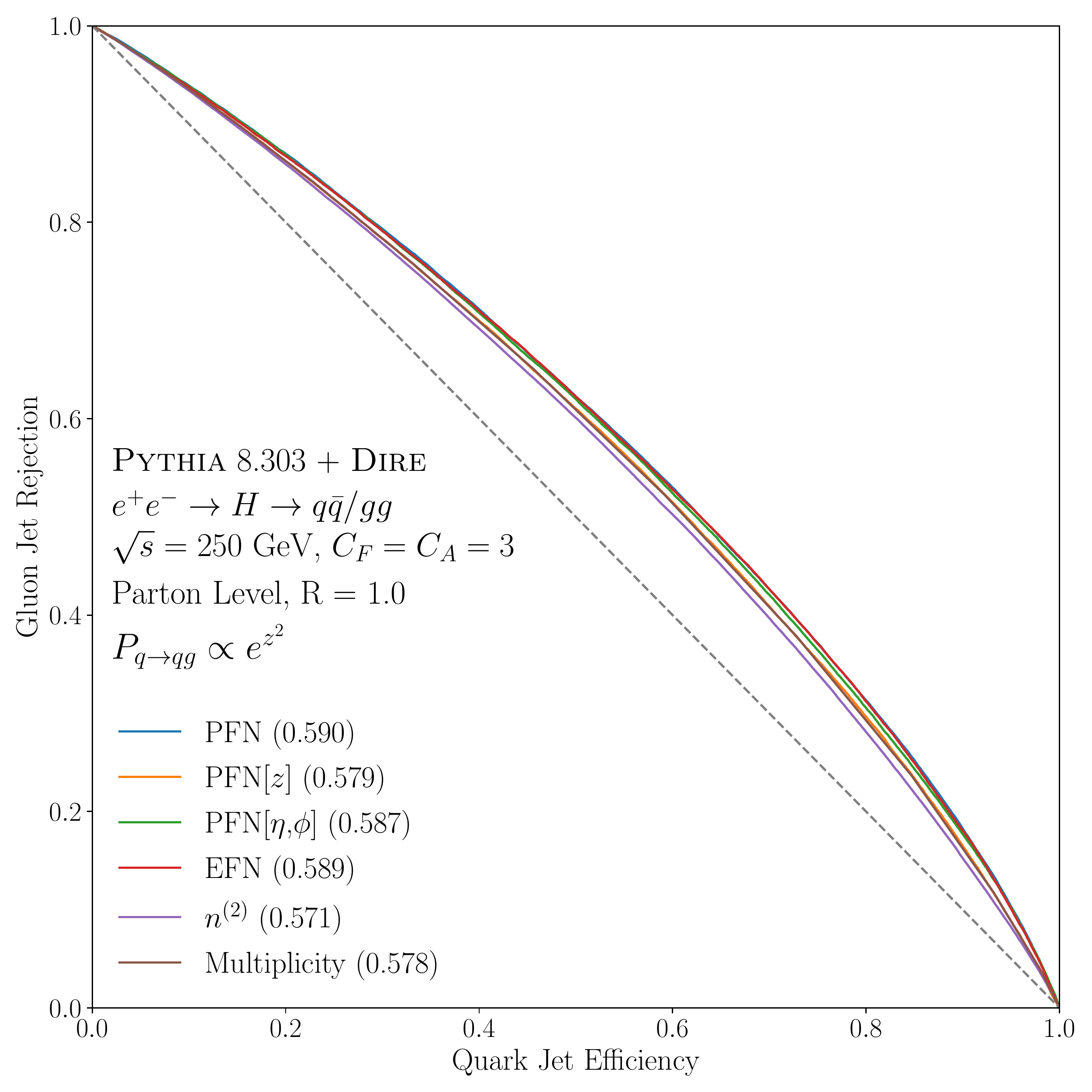}
	\caption{ROC curves and AUC scores (listed in the legend) for the likelihood ratio observable $n^{(2)}$, multiplicity, and the four benchmark DNNs with settings $C_F = C_A = 3$ and $P_{q\to qg} \propto \exp(z^2)$. In the $R = 0.4$ case (left), classifiers that use angular information appear to outperform those using only momenta. This effect mostly vanishes at $R = 1.0$, where all perform approximately equally. In both cases, $n^{(2)}$ performs as well as multiplicity.}
	\label{fig:esq0001}
\end{figure}

Figure \ref{fig:esq0001} shows classifier performance with an $\exp(z^2)$ enhancement to the $q \to qg$ splitting and $C_F = C_A$. The LRO in this case is simply $n^{(2)}$, and it performs as well as multiplicity in both the $R = 0.4$ (left) and $R = 1.0$ (right) cases. There is no \textit{a priori} reason for $n^{(2)}$ to be useful for classification, so these results demonstrate that the LRO does indeed capture useful classification information. Multiplicity, though not explicitly present in $\ln\mathcal{L}$, retains good performance because an enhancement of $q\to qg$ splittings is fundamentally an enhancement of quark jet multiplicity. In both the $R = 0.4$ and $R = 1.0$ cases, multiplicity and $n^{(2)}$ match PFN$[z]$ -- the PFN trained on constituent $z_i$. This indicates that both multiplicity and $n^{(2)}$ capture virtually all of the useful information encoded in the constituent momenta.
\par At $R = 0.4$, we observe a clear performance gap between the computed observables ($n^{(2)}$, multiplicity) and the DNNs that use angular information (PFN, PFN$[\eta,\phi]$, and EFN). This largely disappears at $R = 1.0$, where all classifiers perform within 1-2\% of one another. We speculate that the difference comes from events where the full radiation pattern is not captured within an $R = 0.4$ jet (due to a wide-angle emission, for example). In this case, the computed observables suffer due to incomplete information, while the DNNs exploit potentially spurious angular information. Alternatively, ``behind-the-scenes'' behavior in \textsc{Pythia} -- such as momentum conservation -- may introduce angular correlations that are unaccounted for in the computed observables. The absence of a performance gap at $R = 1.0$ slightly favors the former explanation, and for the remainder of our studies we use $R = 1.0$.

\begin{figure}
	\centering
	\includegraphics[width=0.49\textwidth]{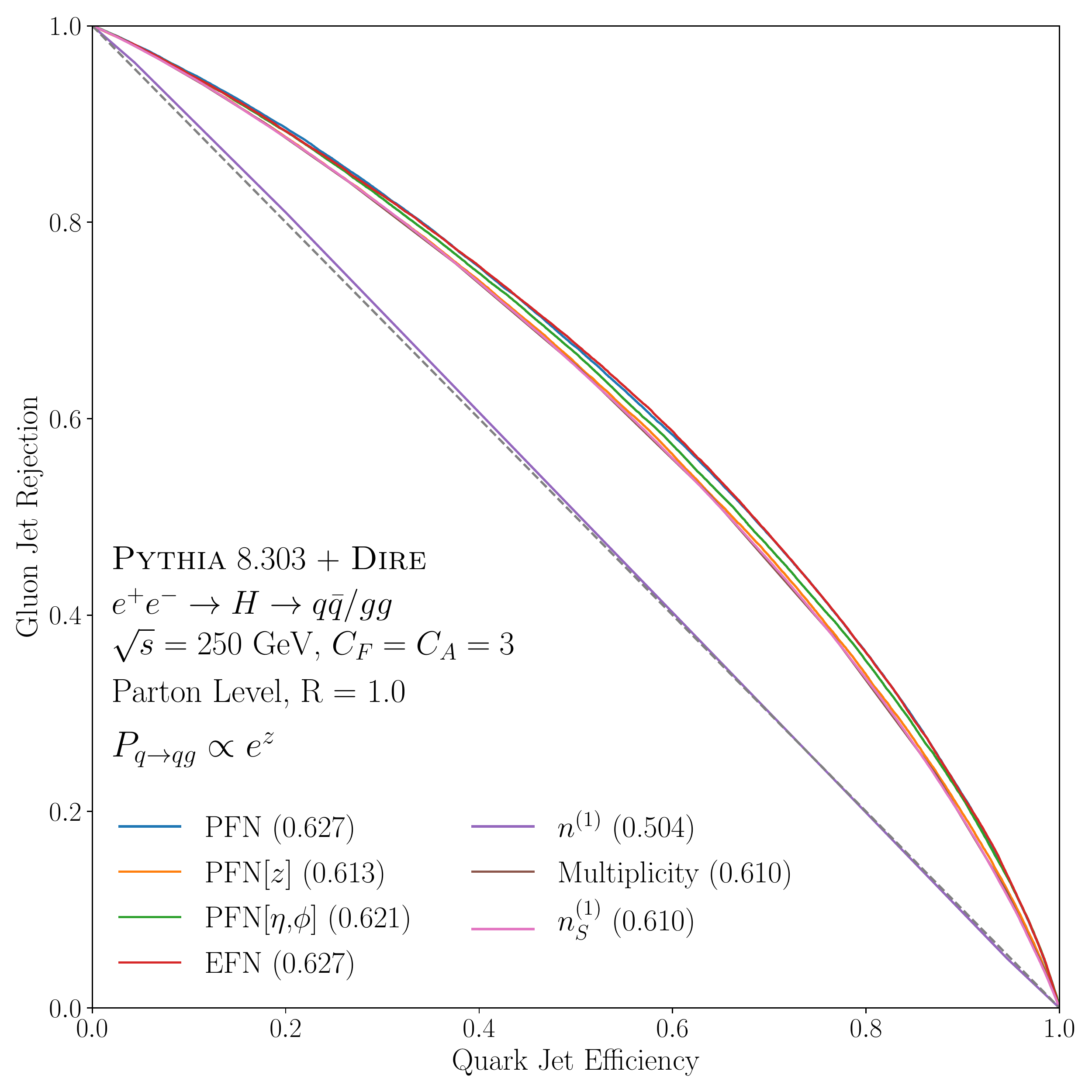}
	\includegraphics[width=0.49\textwidth]{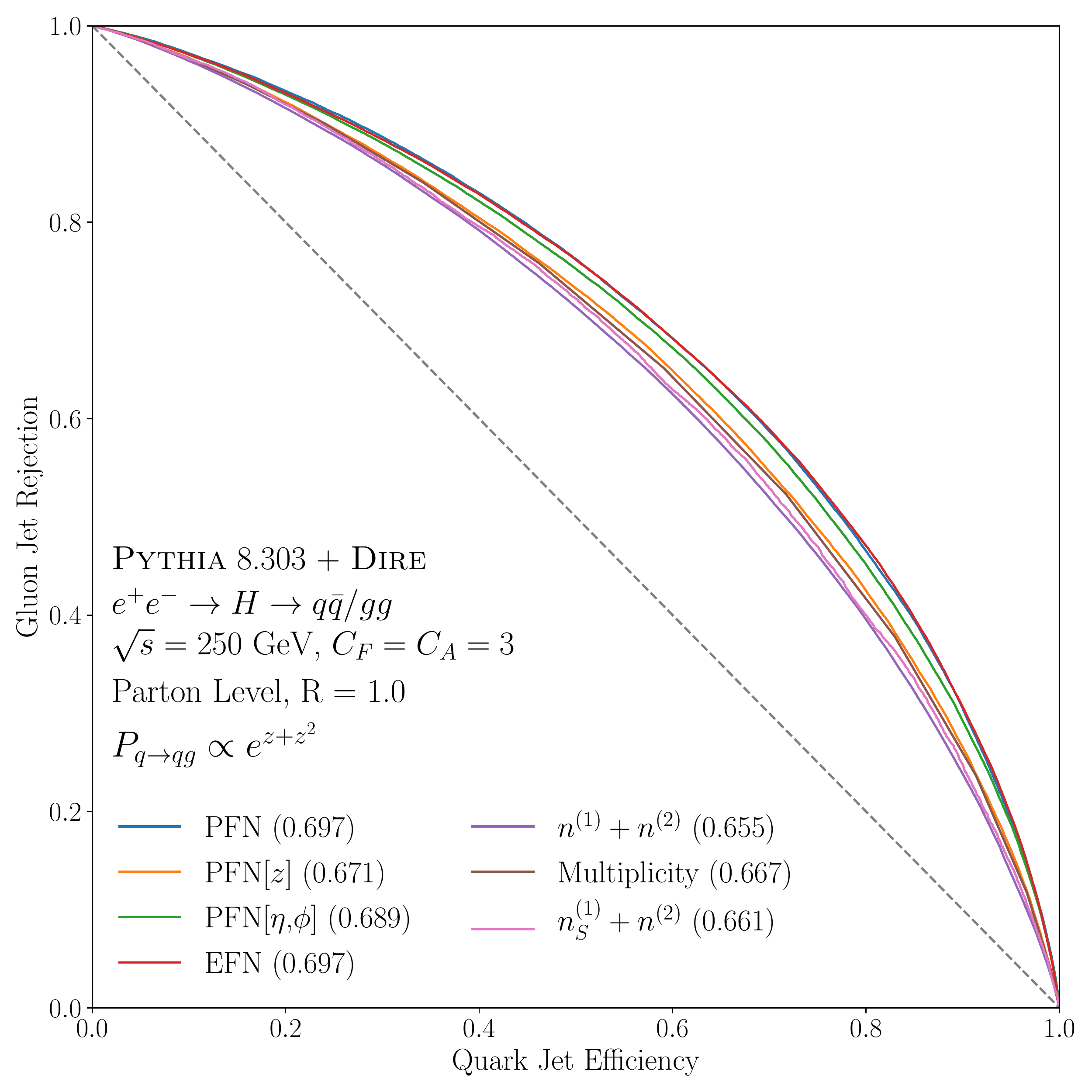}
	\caption{ROC curves and AUCs for the settings $P_{q\to qg} \propto \exp(z)$ (left) and $P_{q\to qg} \propto \exp(z+z^2)$ (right). In each case the likelihood ratio contains the trivial term $n^{(1)}$, and ROCs using alternative formulation $n_S^{(1)}$ of Eq.\ \ref{eq:shannon} are shown in an additional curve.}
	\label{fig:exp_withshan}
\end{figure}

\par Next, we consider the $\exp(z)$ enhancement, which corresponds to the LRO $n^{(1)}$. Unfortunately, $n^{(1)}$ is a nearly the constant unity for all jets and thus useless for discrimination, as shown by the purple curve in Fig.\ \ref{fig:exp_withshan} (left). To circumvent this, we expand the sum $\sum_{i\in \mathrm{jet}} z_i^\kappa$ about $\kappa = 1$, as in Ref.\ \cite{Larkoski:2014pca}. To leading order, this yields:
\begin{equation}
	\label{eq:shannon}
	\sum_{i \in \mathrm{jet}}z_i^\kappa = 1 + \sum_{i\in \mathrm{jet}}(1-\kappa)z_i\ln z_i + \cdots,
\end{equation}
which justifies replacing $n^{(1)}$ with the observable $\sum_{i \in \mathrm{jet}} z_i\ln z_i$. Due to its similarity to Shannon entropy, we denote this observable $n_S^{(1)}$. Figure \ref{fig:exp_withshan} (left) includes the ROC and AUC for $n_S^{(1)}$, which clearly improves upon $n^{(1)}$ and matches the performance of multiplicity and the DNNs, as in the $\exp(z^2)$ case. This again indicates that the LRO captures nearly the same information as the DNNs, while having the clear advantage of interpretability, algebraic simplicity, and ease of computation. 
\par Concluding our studies with $C_F = C_A = 3$, we show results from with $P_{q\to qg} \propto \exp(z+z^2)$ in Fig.\ \ref{fig:exp_withshan} (right). We again include a ROC curve for the corrected LRO $n_{S}^{(1)} + n^{(2)}$, which boosts performance relative to variant using $n^{(1)}$. In this case, however, there is less parity between the computed observables and the DNNs. The PFN, PFN$[\eta,\phi]$, and EFN perform approximately 2\% better than multiplicity and $n_{S}^{(1)} + n^{(2)}$, implying that the angular structure is helpful for classification (though this may be an artifact of the event generator). The LRO is only 1\% behind PFN$[z]$, indicating that it is still a powerful classifier using only the constituent momenta. 

\subsubsection{Restoring $C_F = 4/3$}
\label{sec:cf43}
\begin{figure}[t]
	\centering
	\includegraphics[align=c,width=0.49\textwidth]{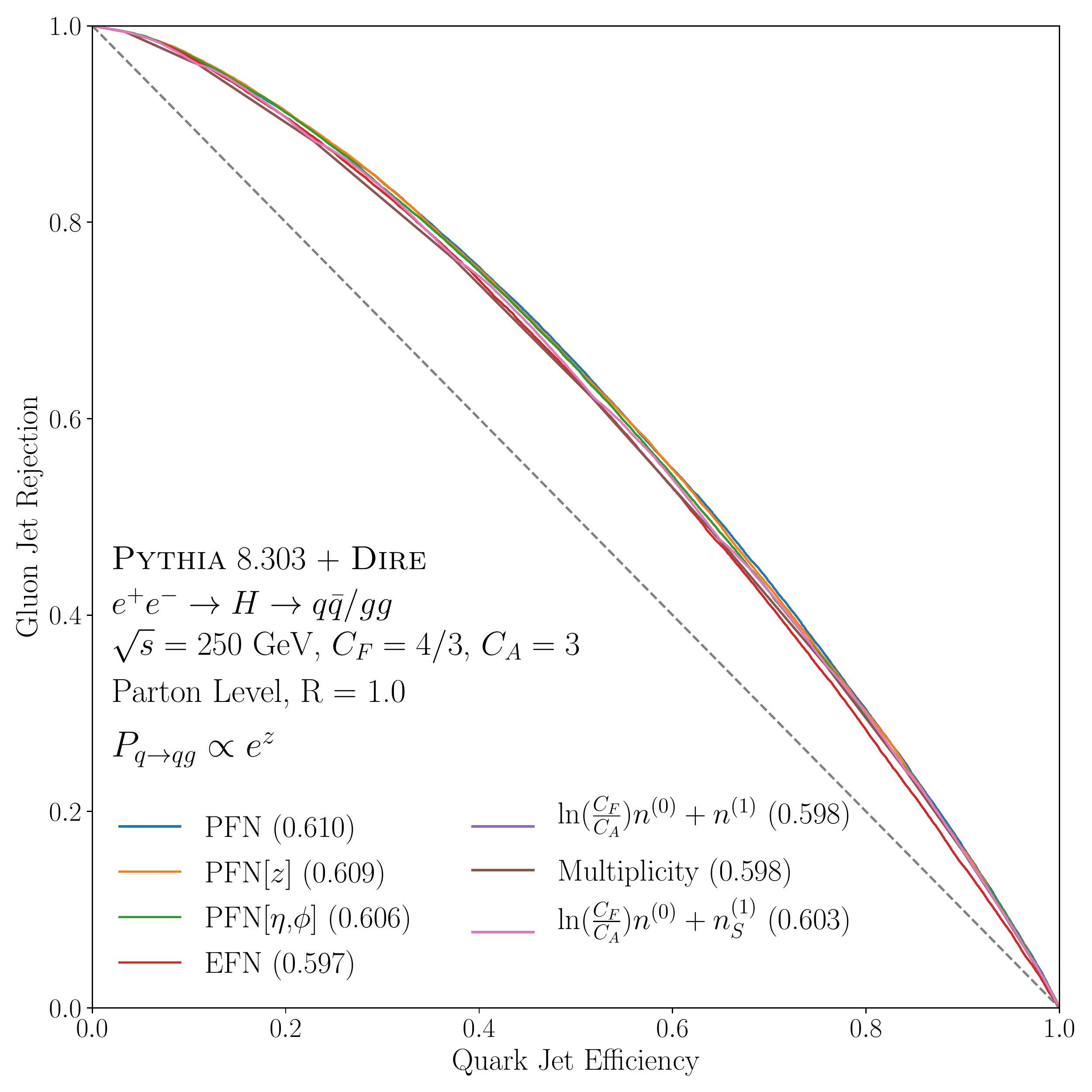}
	\includegraphics[align=c,width=0.49\textwidth]{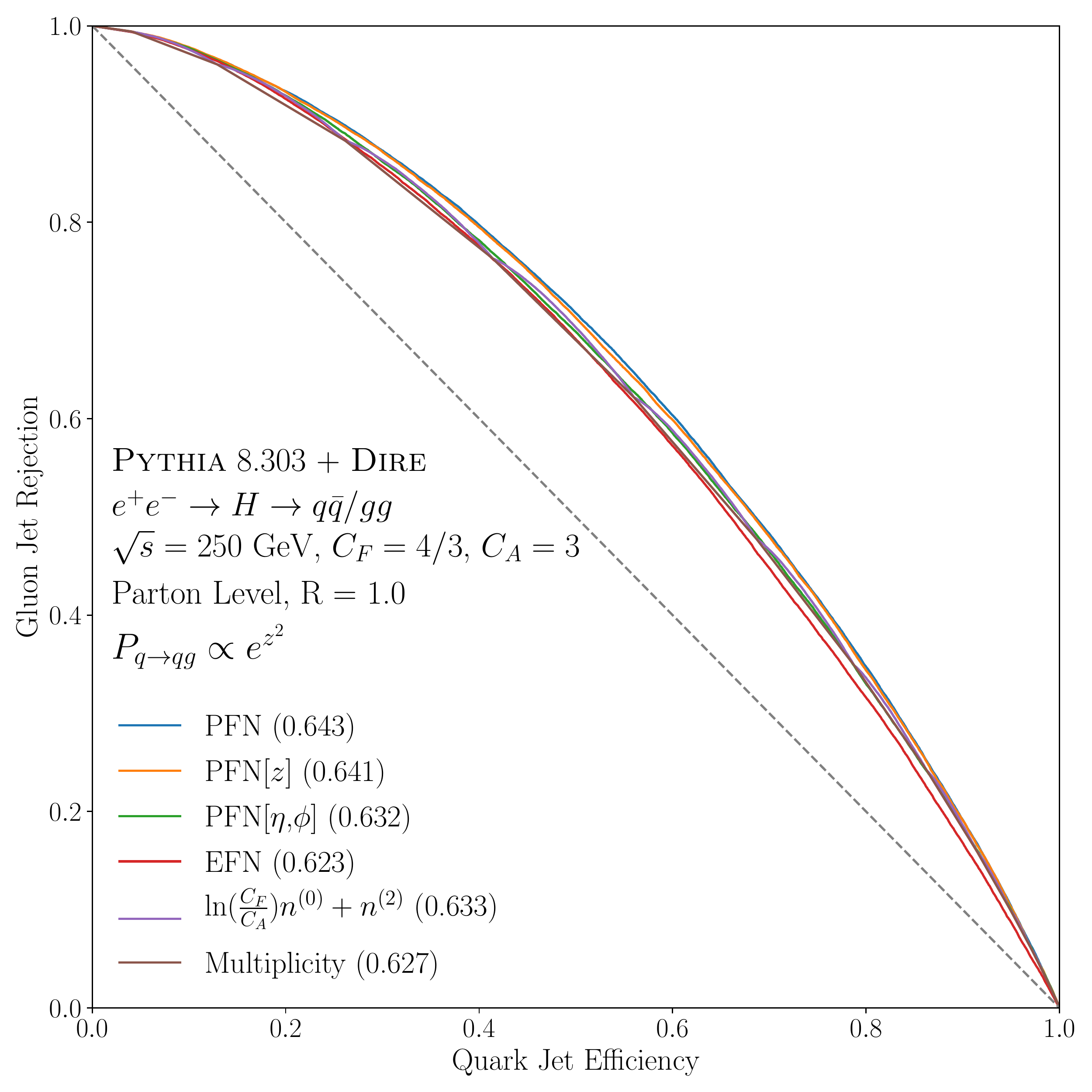} \\
	\includegraphics[align=c,width=0.49\textwidth]{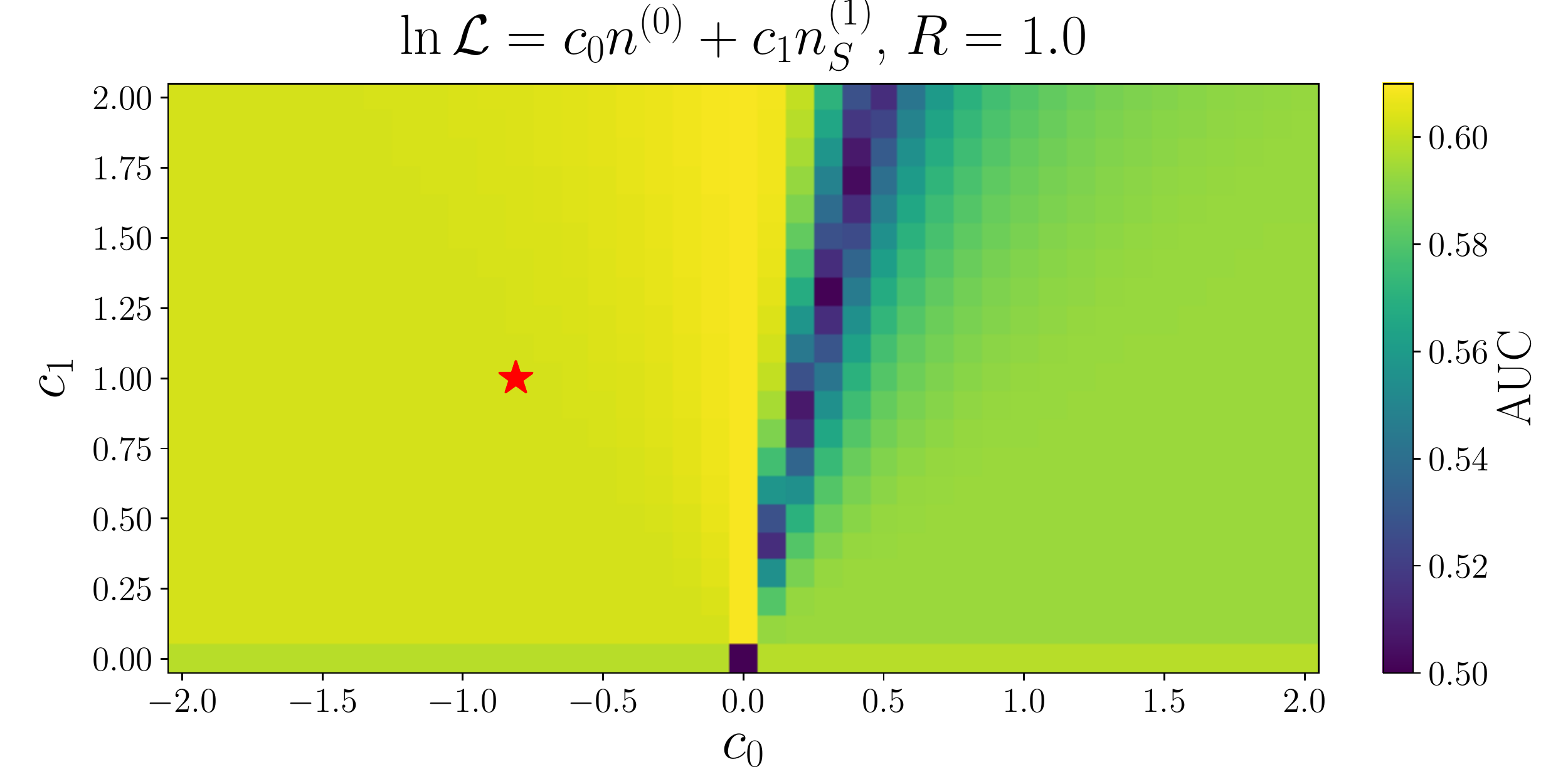} 
	\includegraphics[align=c,width=0.49\textwidth]{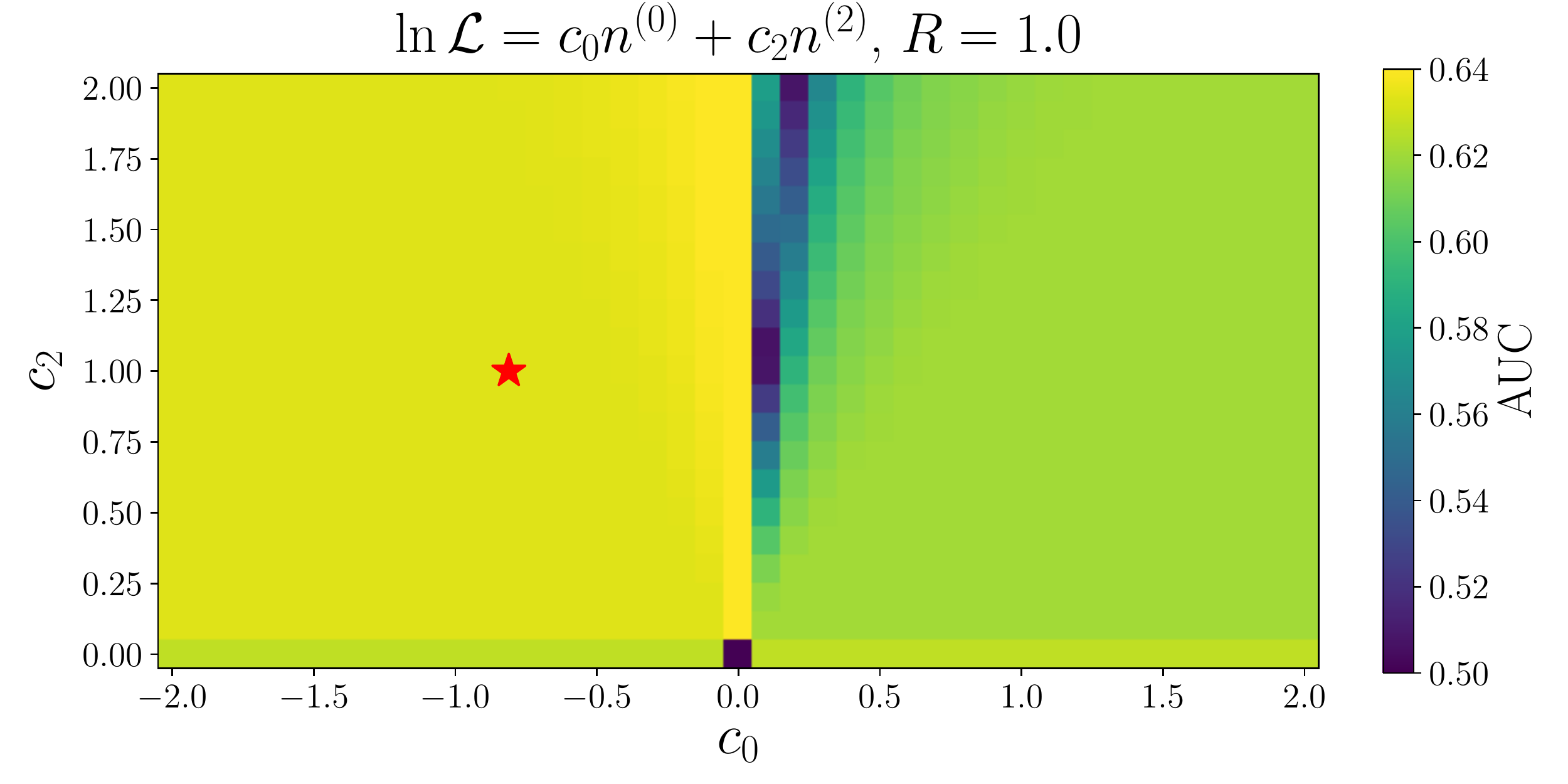}
	\caption{Top: ROC curves and AUC scores for $P_{q\to qg} \propto \exp(z)$ (left) and $P_{q\to qg}\propto \exp(z^2)$ (right), generated with the true quark color factor $C_F = 4/3$. Bottom: AUC scores for a scan over the coefficients $c_0$ and $c_1$/$c_2$ in the observables $c_0n^{(0)} + c_1n_S^{(1)}$ (left) and $c_0n^{(0)} + c_1n^{(2)}$ (right). Color indicates the AUC score, and the grid point represents the $(c_0,c_i)$ setting. In each plot, the red star represents the default working point $c_0 = \ln(C_F/C_A)$, $c_i = 1$ defined by the computed likelihood ratio.}
	\label{fig:CF43}
\end{figure}

To conclude our studies in \textsc{Pythia}, we restore the physical quark color factor $C_F = 4/3$, which introduces an explicit multiplicity term $\ln(C_F/C_A)n^{(0)}$ in the LRO (taking $c_0 = 1$ as mentioned above). Figure \ref{fig:CF43} (top), shows results for $P_{q\to qg} \propto \exp(z)$ (left) and $P_{q\to qg} \propto \exp(z^2)$ (right). The LRO remains very close to the other benchmarks, again indicating that it performs nearly optimally. Notably, the EFN performs worse than the other classifiers, reflecting its insensitivity to jet multiplicity (the leading order effect in the LRO).
\par Since the LRO explicitly contains multiplicity, we are better poised to understand the relative impact of the higher order terms $n_S^{(1)}$ and $n^{(2)}$. In Fig.\ \ref{fig:CF43} (bottom), we scan over the coefficients $(c_0,c_1)$ for the observable $c_0n^{(0)} + c_1n_S^{(1)}$ (left) and $(c_0,c_2)$ for $c_0n^{(0)} + c_2n^{(2)}$ (right).\footnote{For simplicity, have converted the term $\ln(c_0C_F/C_A)n^{(0)}$ into a single prefactor $c_0n^{(0)}$.} We consider points $c_0 \in [-2,2]$ and $c_1,c_2 \in [0,2]$, where negative values of $c_{i>0}$ are omitted because the observable $(c_0,-|c_i|)$ is equivalent to $(-c_0,|c_i|)$ up to an overall sign. At each point, we compute an AUC score for the corresponding LRO and color the points accordingly. The red stars correspond to the default working points $c_0 = \ln(C_F/C_A)$ and $c_1 = c_2 = 1$. The predominant effect appears to be the relative signs of $c_0$ and $c_{i>0}$, with points in the left half (same relative sign as the predicted observable) performing about 1\% better than those on the right half (opposite relative sign). This difference is small, but the performance is very consistent within each half.
\par The other notable feature is the behavior near the lines $c_0 = 0$ (middle vertical line) and $c_{i>0} = 0$ (bottom horizontal line). The points with $c_{i>0} = 0$ -- which lack a contribution from the higher order terms $n_S^{(1)}$ or $n^{(2)}$ -- slightly underperform relative to the points in the upper left half, though the difference is only around 0.5\% in each case. Interestingly, the points that contain \textit{only} higher order terms ($c_0 = 0$) perform the best overall. The differences are again quite small (0.5-1\%) relative to the default point, but slightly more significant (1-1.5\%) relative to the $c_{i>0} = 0$ points. It is curious that the higher order terms alone perform so well, but this is not entirely unexpected as they are somewhat correlated with jet multiplicity. For example, large-multiplicity jets will have many particles with smaller $z_i$, which pushes down the average value of $n^{(2)}$ (a similar trend also holds for $n_S^{(1)}$). Lastly, each plot features a line of poorly performing points emanating from the origin into the right half. These correspond a family of linearly related observables $(rc_0^\prime,rc_{i>0}^\prime)$ ($r \in \mathbb{R}$) for which quark and gluon jets are identically distributed. On either side of this line (i.e.\ the left and right halves of the grid), the distributions differ and enable better discrimination power.
\par As in Sec.\ \ref{sec:expscan}, these trends suggest that the LROs are a good approximation of the true likelihood ratio. The higher order terms clearly encode useful classification information beyond jet multiplicity, and the importance of the relative sign between $n^{(0)}$ and $n_S^{(1)}$ or $n^{(2)}$ reflects the algebraic form of the LROs. Furthermore, the LROs perform within 1\% of the best performing DNN classifiers, indicating that they have captured nearly all of the useful classification information in a simple, computationally tractable expression.

\subsection{Further Investigation: Toy Showers and Hadronization}
\label{sec:duffPS}
\begin{figure}
    \centering
    \includegraphics[width=0.49\textwidth]{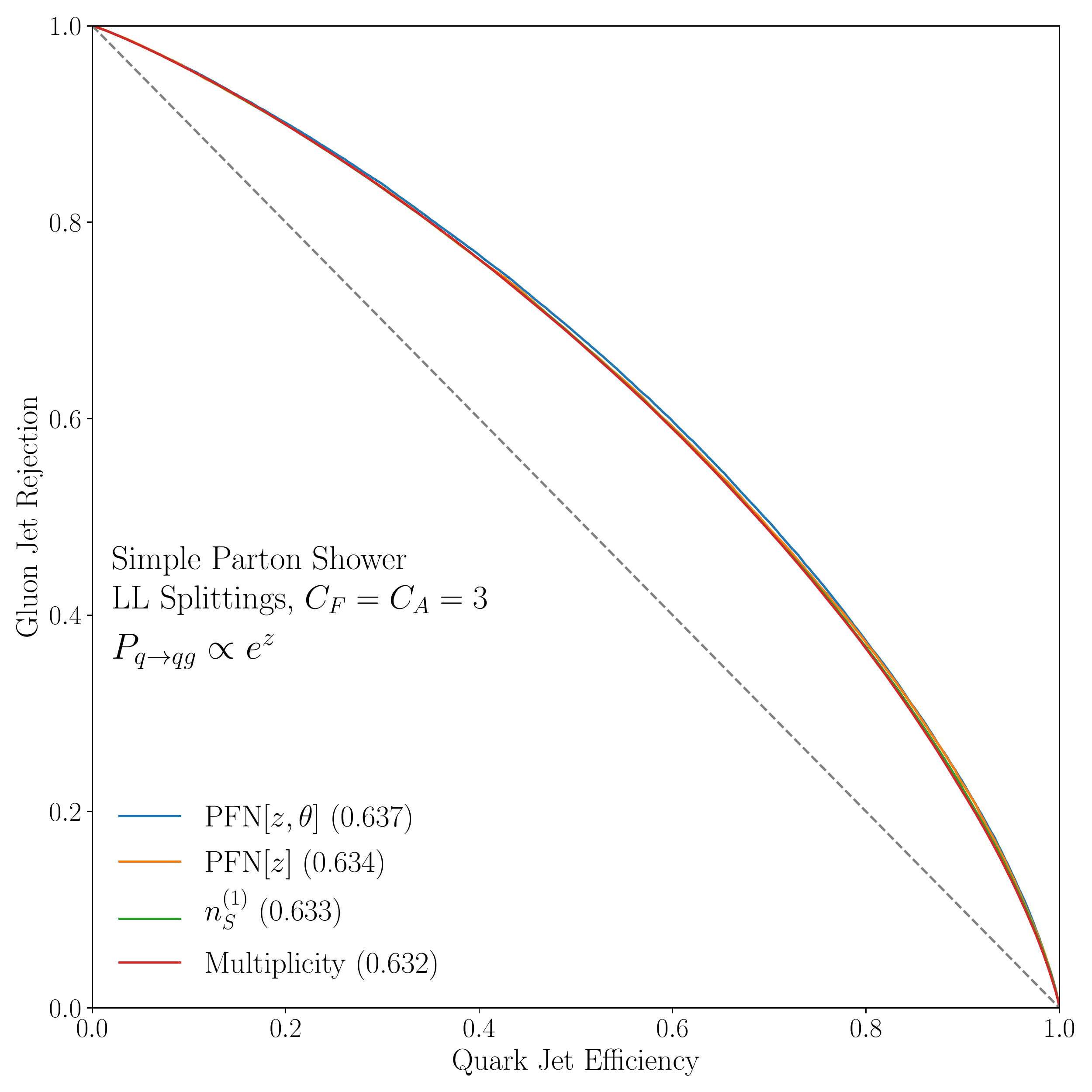}
    \includegraphics[width=0.49\textwidth]{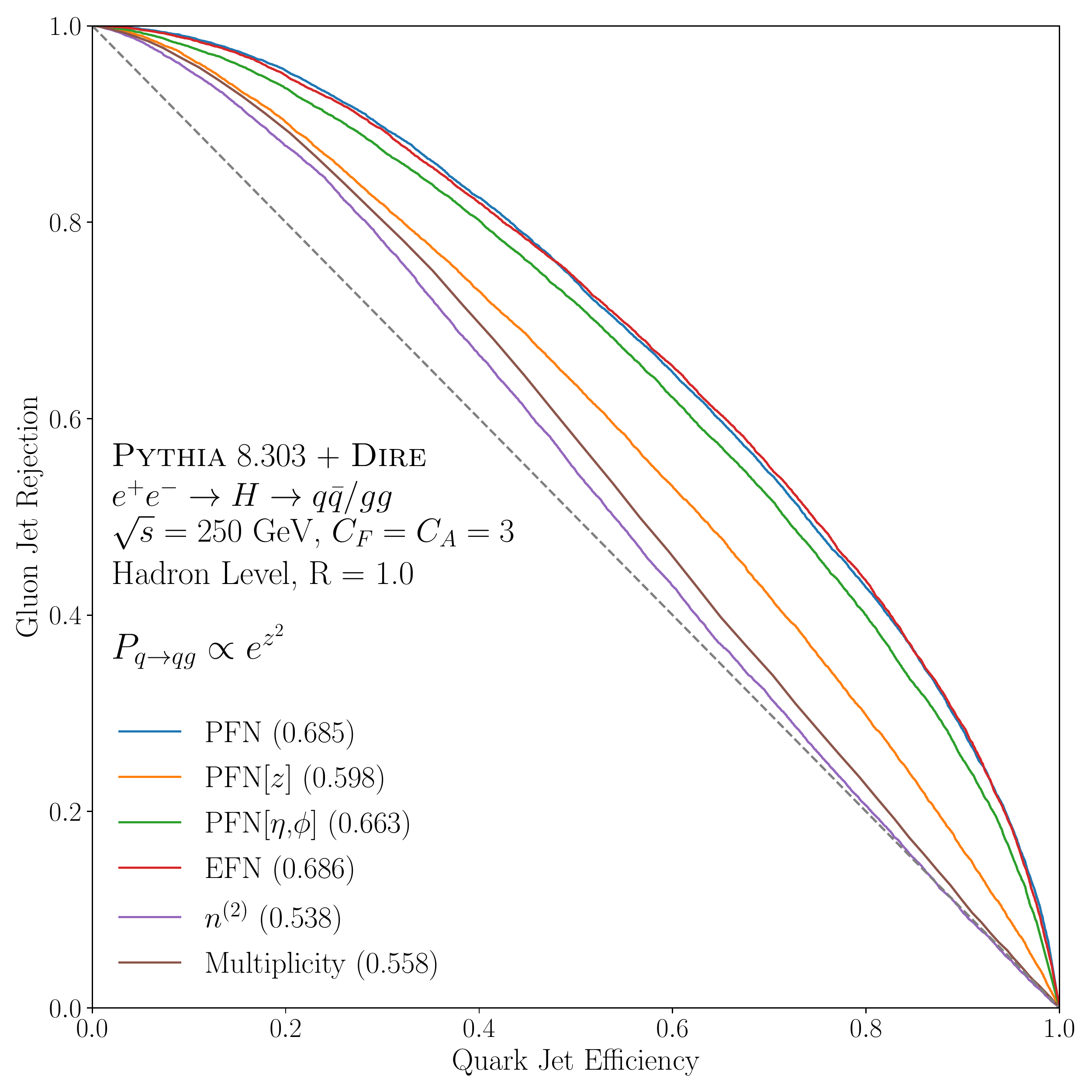}
    \caption{ROC curves comparing predicted classifier performance to neural networks for two additional event generation paradigms: a dedicated jet generator with explicitly tunable splitting functions (left), and \textsc{Pythia} at hadron-level (right). We observe near-perfect agreement between the DNNs and the likelihood ratio observable in the former case. In the latter case, generator-level hadronization effects severely diminish the performance of the calculated observables.}
    \label{fig:duffPS}
\end{figure}
As a final test of the LROs, we examine their performance in classifying jets generated at two opposite extremes: precise adherence to user-specified splitting functions in a simplified parton shower generator, and \textsc{Pythia}/\textsc{DIRE} at hadron-level. In the former case, we expect strong agreement between LROs and DNNs. The simplified generator\footnote{We thank Duff Neill for sharing his parton shower code.} computes splitting probabilities directly from hand-coded likelihood functions $P_{q\to qg}$ and $P_{g \to gg}$, making the correspondence to Eq.\ \ref{eq:diresplits} explicit and the likelihood ratio formulated in Eq.\ \ref{MCObs} nearly exact. In the latter case, the \textsc{Pythia} hadronization routine significantly complicates the correspondence between a parton shower and a final-state jet, and the likelihood ratio formulae we derive no longer apply. We thus expect the LRO performance to drop substantially.
\par Figure \ref{fig:duffPS} shows the results in each scenario. In the simplified scenario (left), there is virtually no difference in performance (less than 0.5\%) between the LRO and the DNNs. The parton shower routine used for these results is essentially a ``jet generator'', and without any of the additional physics features of \textsc{Pythia}/\textsc{DIRE} it allows an even more direct probe of the LROs than the parton-level studies in the previous section. While this result is unsurprising, it is strong evidence that the LROs -- where they apply nearly exactly -- do in fact capture the same information as a DNN. In the case with full hadronization (right), the LRO loses nearly all of its discrimination power, performing worse than multiplicity and all of the DNNs. The performance drop is expected, but the magnitude is exceptionally large and underscores the limitations of working at parton-level when constructing the LROs.

\subsection{Discussion}
In each result shown in Secs.\ \ref{sec:expscan} and \ref{sec:cf43}, our predicted LROs track closely with jet multiplicity and PFN$[z]$, and in most cases perform within 1 to 2\% of the full PFN. Assuming that PFN$[z]$ learns the optimal classifier given the constituent momenta, this implies that the LROs are near optimal and roughly equivalent to the true likelihood ratio. The overall parity with multiplicity suggests that while LROs may not confer an explicit \textit{advantage}, they capture the same information and can be systematically computed to one's desired accuracy. In the $C_F \neq C_A$ case -- where multiplicity is explicitly included in the LRO -- the higher order terms do appear to confer a small advantage. In the hyper-simplified case addressed in Sec.\ \ref{sec:duffPS}, we see very clear evidence that the LRO -- when it reflects the true ``physics'' in the event generator -- does classify optimally. Broadly speaking, it is encouraging that the LROs perform so well relative to multiplicity and DNNs on \textsc{Pythia} jets, given that they are (a) calculated using several simplifying assumptions, and (b) insensitive to any angular substructure created by the generator. These results demonstrate the value of an approximate likelihood ratio when the true ratio is intractable, and motivate further study of physics-motivated observables that can approach DNN performance in more realistic scenarios.

%%%%%%%%%%%%%%%%%%%
\section{Conclusions}
\label{sec:conc}
%%%%%%%%%%%%%%%%%%%

In this paper we have presented a new approach to understanding jet substructure observables, by analyzing their likelihood function at a fixed accuracy and explicitly constructing likelihood ratios. In the case of quark/gluon jet discrimination, we have proven that multiplicity is optimal at leading order, and that going to higher perturbative order introduces sub-leading contributions in the form of weighted multiplicities. Furthermore, we have shown that for observables that can be computed using eikonal splitting functions, their LL result cannot beat multiplicity. Beyond its application in quark/gluon jet classification, our technique reveals a general strategy for computing and systematically improving approximations of the likelihood ratio for a wide range of observables.

We demonstrate our approach using Monte Carlo simulations of parton-level jets, and directly compare the analytic classifier to a collection of deep neural networks. Using jets clustered from fixed-order parton showers, we find that the analytic classifiers nearly match the performance of Particle Flow Networks. This indicates that the networks indeed ``discover" the underlying structure of the shower and extract relevant classification information. In some cases, the networks retain a clear advantage which may be attributed to artifacts of the jet clustering scheme, quirks of the event generator, or constraints that are unaccounted for in our calculations (e.g.\ momentum conservation).\footnote{A detailed analysis that explicitly considers jet clustering can be found in Ref.~\cite{Dreyer:2021hhr}.} In a toy parton shower where splitting functions are hand-coded and the full shower final state is recorded, the analytic and PFN classifiers perform identically.

In summary, this work demonstrates the possibility of a first-principles physical explanation for the performance of ML classifiers. This signals a new way for theory and ML to work together, wherein ideas from statistics and deep learning (e.g.\ the optimality of likelihood ratios) motivate theoretical calculations that can at least partially explain the performance of a deep neural network. While it is unlikely that analytic classifiers can completely replace DNNs, especially in the noisy and high-dimensional space of collider data, they can illuminate some corners of the ``black box''. Furthermore, they can enable an understanding of which physical features of parton shower simulations are being exploited by classifiers, to ensure that they are well described. 

We look forward to continued progress in this vein, and believe that our approach is general enough to apply beyond quark/gluon discrimination. In future work, it would be interesting to modify the technique to become more robust against unconstrained behaviors of Monte Carlo generators, or apply it to other classification problems such as boosted $W/Z/h$ boson identification.  

\acknowledgments

We are grateful to Eric~Metodiev for collaboration during an early stage of this work, including the development of many components of the codebase. %, and for feedback on the manuscript.

The authors are grateful to Gregory Soyez, and Jesse Thaler for helpful discussions and suggestions. We thank Duff Neill for sharing his parton shower code with us, and for useful discussions.
%
%The work of E.M.M. is supported by the Office of Nuclear Physics of the U.S. Department of Energy (DOE) under grant DE-SC0011090 and the DOE Office of High Energy Physics under grant DE-SC-00012567.
%
The work of B.N. is supported by the DOE Office of Science under contract DE-AC02-05CH11231.
Cloud computing resources were provided through a Microsoft Azure for Research award.
I.M. is supported by start-up funds from Yale University.

\bibliographystyle{jhep}
\bibliography{myrefs}

\providecommand{\href}[2]{#2}\begingroup\raggedright\begin{thebibliography}{10}

\bibitem{Abdesselam:2010pt}
A.~Abdesselam et~al., {\it {Boosted objects: A Probe of beyond the Standard
  Model physics}},  {\em Eur. Phys. J.} {\bf C71} (2011) 1661,
  [\href{http://arxiv.org/abs/1012.5412}{{\tt arXiv:1012.5412}}].

\bibitem{Altheimer:2012mn}
A.~Altheimer et~al., {\it {Jet Substructure at the Tevatron and LHC: New
  results, new tools, new benchmarks}},  {\em J. Phys.} {\bf G39} (2012)
  063001, [\href{http://arxiv.org/abs/1201.0008}{{\tt arXiv:1201.0008}}].

\bibitem{Altheimer:2013yza}
A.~Altheimer et~al., {\it {Boosted objects and jet substructure at the LHC.
  Report of BOOST2012, held at IFIC Valencia, 23rd-27th of July 2012}},  {\em
  Eur. Phys. J.} {\bf C74} (2014), no.~3 2792,
  [\href{http://arxiv.org/abs/1311.2708}{{\tt arXiv:1311.2708}}].

\bibitem{Adams:2015hiv}
D.~Adams et~al., {\it {Towards an Understanding of the Correlations in Jet
  Substructure}},  {\em Eur. Phys. J.} {\bf C75} (2015), no.~9 409,
  [\href{http://arxiv.org/abs/1504.00679}{{\tt arXiv:1504.00679}}].

\bibitem{Larkoski:2017jix}
A.~J. Larkoski, I.~Moult, and B.~Nachman, {\it {Jet Substructure at the Large
  Hadron Collider: A Review of Recent Advances in Theory and Machine
  Learning}},  \href{http://arxiv.org/abs/1709.04464}{{\tt arXiv:1709.04464}}.

\bibitem{Kogler:2018hem}
R.~Kogler et~al., {\it {Jet Substructure at the Large Hadron Collider:
  Experimental Review}},  {\em Rev. Mod. Phys.} {\bf 91} (2019), no.~4 045003,
  [\href{http://arxiv.org/abs/1803.06991}{{\tt arXiv:1803.06991}}].

\bibitem{Marzani:2019hun}
S.~Marzani, G.~Soyez, and M.~Spannowsky, {\em {Looking inside jets: an
  introduction to jet substructure and boosted-object phenomenology}},
  vol.~958.
\newblock Springer, 2019.

\bibitem{Larkoski:2013paa}
A.~J. Larkoski and J.~Thaler, {\it {Unsafe but Calculable: Ratios of
  Angularities in Perturbative QCD}},  {\em JHEP} {\bf 09} (2013) 137,
  [\href{http://arxiv.org/abs/1307.1699}{{\tt arXiv:1307.1699}}].

\bibitem{Larkoski:2015lea}
A.~J. Larkoski, S.~Marzani, and J.~Thaler, {\it {Sudakov Safety in Perturbative
  QCD}},  {\em Phys. Rev. D} {\bf 91} (2015), no.~11 111501,
  [\href{http://arxiv.org/abs/1502.01719}{{\tt arXiv:1502.01719}}].

\bibitem{Larkoski:2014gra}
A.~J. Larkoski, I.~Moult, and D.~Neill, {\it {Power Counting to Better Jet
  Observables}},  {\em JHEP} {\bf 12} (2014) 009,
  [\href{http://arxiv.org/abs/1409.6298}{{\tt arXiv:1409.6298}}].

\bibitem{Larkoski:2014zma}
A.~J. Larkoski, I.~Moult, and D.~Neill, {\it {Building a Better Boosted Top
  Tagger}},  {\em Phys. Rev. D} {\bf 91} (2015), no.~3 034035,
  [\href{http://arxiv.org/abs/1411.0665}{{\tt arXiv:1411.0665}}].

\bibitem{Larkoski:2015kga}
A.~J. Larkoski, I.~Moult, and D.~Neill, {\it {Analytic Boosted Boson
  Discrimination}},  {\em JHEP} {\bf 05} (2016) 117,
  [\href{http://arxiv.org/abs/1507.03018}{{\tt arXiv:1507.03018}}].

\bibitem{Moult:2016cvt}
I.~Moult, L.~Necib, and J.~Thaler, {\it {New Angles on Energy Correlation
  Functions}},  {\em JHEP} {\bf 12} (2016) 153,
  [\href{http://arxiv.org/abs/1609.07483}{{\tt arXiv:1609.07483}}].

\bibitem{Larkoski:2017iuy}
A.~J. Larkoski, I.~Moult, and D.~Neill, {\it {Analytic Boosted Boson
  Discrimination at the Large Hadron Collider}},
  \href{http://arxiv.org/abs/1708.06760}{{\tt arXiv:1708.06760}}.

\bibitem{Larkoski:2017cqq}
A.~J. Larkoski, I.~Moult, and D.~Neill, {\it {Factorization and Resummation for
  Groomed Multi-Prong Jet Shapes}},  {\em JHEP} {\bf 02} (2018) 144,
  [\href{http://arxiv.org/abs/1710.00014}{{\tt arXiv:1710.00014}}].

\bibitem{Cal:2022fnm}
P.~Cal, J.~Thaler, and W.~J. Waalewijn, {\it {Power Counting Energy Flow
  Polynomials}},  \href{http://arxiv.org/abs/2205.06818}{{\tt
  arXiv:2205.06818}}.

\bibitem{Gellersen:2021eci}
L.~Gellersen, S.~H\"oche, and S.~Prestel, {\it {Disentangling soft and
  collinear effects in QCD parton showers}},  {\em Phys. Rev. D} {\bf 105}
  (2022), no.~11 114012, [\href{http://arxiv.org/abs/2110.05964}{{\tt
  arXiv:2110.05964}}].

\bibitem{Dulat:2018vuy}
F.~Dulat, S.~H\"oche, and S.~Prestel, {\it {Leading-Color Fully Differential
  Two-Loop Soft Corrections to QCD Dipole Showers}},  {\em Phys. Rev. D} {\bf
  98} (2018), no.~7 074013, [\href{http://arxiv.org/abs/1805.03757}{{\tt
  arXiv:1805.03757}}].

\bibitem{Hoche:2017hno}
S.~H\"oche, F.~Krauss, and S.~Prestel, {\it {Implementing NLO DGLAP evolution
  in Parton Showers}},  {\em JHEP} {\bf 10} (2017) 093,
  [\href{http://arxiv.org/abs/1705.00982}{{\tt arXiv:1705.00982}}].

\bibitem{Li:2016yez}
H.~T. Li and P.~Skands, {\it {A framework for second-order parton showers}},
  {\em Phys. Lett. B} {\bf 771} (2017) 59--66,
  [\href{http://arxiv.org/abs/1611.00013}{{\tt arXiv:1611.00013}}].

\bibitem{Hoche:2017iem}
S.~H\"oche and S.~Prestel, {\it {Triple collinear emissions in parton
  showers}},  {\em Phys. Rev. D} {\bf 96} (2017), no.~7 074017,
  [\href{http://arxiv.org/abs/1705.00742}{{\tt arXiv:1705.00742}}].

\bibitem{vanBeekveld:2022zhl}
M.~van Beekveld, S.~Ferrario~Ravasio, G.~P. Salam, A.~Soto-Ontoso, G.~Soyez,
  and R.~Verheyen, {\it {PanScales parton showers for hadron collisions:
  formulation and fixed-order studies}},
  \href{http://arxiv.org/abs/2205.02237}{{\tt arXiv:2205.02237}}.

\bibitem{Hamilton:2021dyz}
K.~Hamilton, A.~Karlberg, G.~P. Salam, L.~Scyboz, and R.~Verheyen, {\it {Soft
  spin correlations in final-state parton showers}},  {\em JHEP} {\bf 03}
  (2022) 193, [\href{http://arxiv.org/abs/2111.01161}{{\tt arXiv:2111.01161}}].

\bibitem{Karlberg:2021kwr}
A.~Karlberg, G.~P. Salam, L.~Scyboz, and R.~Verheyen, {\it {Spin correlations
  in final-state parton showers and jet observables}},  {\em Eur. Phys. J. C}
  {\bf 81} (2021), no.~8 681, [\href{http://arxiv.org/abs/2103.16526}{{\tt
  arXiv:2103.16526}}].

\bibitem{Hamilton:2020rcu}
K.~Hamilton, R.~Medves, G.~P. Salam, L.~Scyboz, and G.~Soyez, {\it {Colour and
  logarithmic accuracy in final-state parton showers}},  {\em JHEP} {\bf 03}
  (2021), no.~041 041, [\href{http://arxiv.org/abs/2011.10054}{{\tt
  arXiv:2011.10054}}].

\bibitem{Dasgupta:2020fwr}
M.~Dasgupta, F.~A. Dreyer, K.~Hamilton, P.~F. Monni, G.~P. Salam, and G.~Soyez,
  {\it {Parton showers beyond leading logarithmic accuracy}},  {\em Phys. Rev.
  Lett.} {\bf 125} (2020), no.~5 052002,
  [\href{http://arxiv.org/abs/2002.11114}{{\tt arXiv:2002.11114}}].

\bibitem{neyman1933ix}
J.~Neyman and E.~S. Pearson, {\it Ix. on the problem of the most efficient
  tests of statistical hypotheses},  {\em Phil. Trans. R. Soc. Lond. A} {\bf
  231} (1933), no.~694-706 289--337.

\bibitem{D0:2004rvt}
{\bf D0} Collaboration, V.~M. Abazov et~al., {\it {A precision measurement of
  the mass of the top quark}},  {\em Nature} {\bf 429} (2004) 638--642,
  [\href{http://arxiv.org/abs/hep-ex/0406031}{{\tt hep-ex/0406031}}].

\bibitem{Soper:2011cr}
D.~E. Soper and M.~Spannowsky, {\it {Finding physics signals with shower
  deconstruction}},  {\em Phys. Rev. D} {\bf 84} (2011) 074002,
  [\href{http://arxiv.org/abs/1102.3480}{{\tt arXiv:1102.3480}}].

\bibitem{Soper:2012pb}
D.~E. Soper and M.~Spannowsky, {\it {Finding top quarks with shower
  deconstruction}},  {\em Phys. Rev. D} {\bf 87} (2013) 054012,
  [\href{http://arxiv.org/abs/1211.3140}{{\tt arXiv:1211.3140}}].

\bibitem{FerreiradeLima:2016gcz}
D.~Ferreira~de Lima, P.~Petrov, D.~Soper, and M.~Spannowsky, {\it {Quark-Gluon
  tagging with Shower Deconstruction: Unearthing dark matter and Higgs
  couplings}},  {\em Phys. Rev. D} {\bf 95} (2017), no.~3 034001,
  [\href{http://arxiv.org/abs/1607.06031}{{\tt arXiv:1607.06031}}].

\bibitem{Larkoski:2019nwj}
A.~J. Larkoski and E.~M. Metodiev, {\it {A Theory of Quark vs. Gluon
  Discrimination}},  {\em JHEP} {\bf 10} (2019) 014,
  [\href{http://arxiv.org/abs/1906.01639}{{\tt arXiv:1906.01639}}].

\bibitem{Kasieczka:2020nyd}
G.~Kasieczka, S.~Marzani, G.~Soyez, and G.~Stagnitto, {\it {Towards Machine
  Learning Analytics for Jet Substructure}},  {\em JHEP} {\bf 09} (2020) 195,
  [\href{http://arxiv.org/abs/2007.04319}{{\tt arXiv:2007.04319}}].

\bibitem{Bieringer:2020tnw}
S.~Bieringer, A.~Butter, T.~Heimel, S.~H\"oche, U.~K\"othe, T.~Plehn, and S.~T.
  Radev, {\it {Measuring QCD Splittings with Invertible Networks}},  {\em
  SciPost Phys.} {\bf 10} (2021), no.~6 126,
  [\href{http://arxiv.org/abs/2012.09873}{{\tt arXiv:2012.09873}}].

\bibitem{Dreyer:2021hhr}
F.~Dreyer, G.~Soyez, and A.~Takacs, {\it {Quarks and gluons in the Lund
  plane}},  \href{http://arxiv.org/abs/2112.09140}{{\tt arXiv:2112.09140}}.

\bibitem{Lai:2020byl}
Y.~S. Lai, D.~Neill, M.~P\l{}osko\'n, and F.~Ringer, {\it {Explainable machine
  learning of the underlying physics of high-energy particle collisions}},
  {\em Phys. Lett. B} {\bf 829} (2022) 137055,
  [\href{http://arxiv.org/abs/2012.06582}{{\tt arXiv:2012.06582}}].

\bibitem{Buckley:2020kdp}
A.~Buckley, G.~Callea, A.~J. Larkoski, and S.~Marzani, {\it {An Optimal
  Observable for Color Singlet Identification}},  {\em SciPost Phys.} {\bf 9}
  (2020) 026, [\href{http://arxiv.org/abs/2006.10480}{{\tt arXiv:2006.10480}}].

\bibitem{Dreyer:2018nbf}
F.~A. Dreyer, G.~P. Salam, and G.~Soyez, {\it {The Lund Jet Plane}},  {\em
  JHEP} {\bf 12} (2018) 064, [\href{http://arxiv.org/abs/1807.04758}{{\tt
  arXiv:1807.04758}}].

\bibitem{Dokshitzer:2005bf}
{\relax Yu}.~L. Dokshitzer, G.~Marchesini, and G.~P. Salam, {\it {Revisiting
  parton evolution and the large-x limit}},  {\em Phys. Lett.} {\bf B634}
  (2006) 504--507, [\href{http://arxiv.org/abs/hep-ph/0511302}{{\tt
  hep-ph/0511302}}].

\bibitem{Dokshitzer:2006nm}
{\relax Yu}.~L. Dokshitzer and G.~Marchesini, {\it {N=4 SUSY Yang-Mills: three
  loops made simple(r)}},  {\em Phys. Lett.} {\bf B646} (2007) 189--201,
  [\href{http://arxiv.org/abs/hep-th/0612248}{{\tt hep-th/0612248}}].

\bibitem{Beccaria:2007bb}
M.~Beccaria, {\relax Yu}.~L. Dokshitzer, and G.~Marchesini, {\it {Twist 3 of
  the sl(2) sector of N=4 SYM and reciprocity respecting evolution}},  {\em
  Phys. Lett.} {\bf B652} (2007) 194--202,
  [\href{http://arxiv.org/abs/0705.2639}{{\tt arXiv:0705.2639}}].

\bibitem{Levy:1969cr}
M.~Levy and J.~Sucher, {\it {Eikonal approximation in quantum field theory}},
  {\em Phys. Rev.} {\bf 186} (1969) 1656--1670.

\bibitem{Frye:2017yrw}
C.~Frye, A.~J. Larkoski, J.~Thaler, and K.~Zhou, {\it {Casimir Meets Poisson:
  Improved Quark/Gluon Discrimination with Counting Observables}},  {\em JHEP}
  {\bf 09} (2017) 083, [\href{http://arxiv.org/abs/1704.06266}{{\tt
  arXiv:1704.06266}}].

\bibitem{Dixon:2019uzg}
L.~J. Dixon, I.~Moult, and H.~X. Zhu, {\it {Collinear limit of the
  energy-energy correlator}},  {\em Phys. Rev. D} {\bf 100} (2019), no.~1
  014009, [\href{http://arxiv.org/abs/1905.01310}{{\tt arXiv:1905.01310}}].

\bibitem{Chen:2019bpb}
H.~Chen, M.-X. Luo, I.~Moult, T.-Z. Yang, X.~Zhang, and H.~X. Zhu, {\it {Three
  point energy correlators in the collinear limit: symmetries, dualities and
  analytic results}},  {\em JHEP} {\bf 08} (2020), no.~08 028,
  [\href{http://arxiv.org/abs/1912.11050}{{\tt arXiv:1912.11050}}].

\bibitem{Chen:2020vvp}
H.~Chen, I.~Moult, X.~Zhang, and H.~X. Zhu, {\it {Rethinking jets with energy
  correlators: Tracks, resummation, and analytic continuation}},  {\em Phys.
  Rev. D} {\bf 102} (2020), no.~5 054012,
  [\href{http://arxiv.org/abs/2004.11381}{{\tt arXiv:2004.11381}}].

\bibitem{Chen:2020adz}
H.~Chen, I.~Moult, and H.~X. Zhu, {\it {Quantum Interference in Jet
  Substructure from Spinning Gluons}},  {\em Phys. Rev. Lett.} {\bf 126}
  (2021), no.~11 112003, [\href{http://arxiv.org/abs/2011.02492}{{\tt
  arXiv:2011.02492}}].

\bibitem{Chen:2021gdk}
H.~Chen, I.~Moult, and H.~X. Zhu, {\it {Spinning Gluons from the QCD Light-Ray
  OPE}},  \href{http://arxiv.org/abs/2104.00009}{{\tt arXiv:2104.00009}}.

\bibitem{Komiske:2022enw}
P.~T. Komiske, I.~Moult, J.~Thaler, and H.~X. Zhu, {\it {Analyzing N-point
  Energy Correlators Inside Jets with CMS Open Data}},
  \href{http://arxiv.org/abs/2201.07800}{{\tt arXiv:2201.07800}}.

\bibitem{Holguin:2022epo}
J.~Holguin, I.~Moult, A.~Pathak, and M.~Procura, {\it {A New Paradigm for
  Precision Top Physics: Weighing the Top with Energy Correlators}},
  \href{http://arxiv.org/abs/2201.08393}{{\tt arXiv:2201.08393}}.

\bibitem{Chen:2022jhb}
H.~Chen, I.~Moult, J.~Sandor, and H.~X. Zhu, {\it {Celestial Blocks and
  Transverse Spin in the Three-Point Energy Correlator}},
  \href{http://arxiv.org/abs/2202.04085}{{\tt arXiv:2202.04085}}.

\bibitem{Chen:2022swd}
H.~Chen, I.~Moult, J.~Thaler, and H.~X. Zhu, {\it {Non-Gaussianities in
  Collider Energy Flux}},  \href{http://arxiv.org/abs/2205.02857}{{\tt
  arXiv:2205.02857}}.

\bibitem{Lee:2022ige}
K.~Lee, B.~Me\c{c}aj, and I.~Moult, {\it {Conformal Colliders Meet the LHC}},
  \href{http://arxiv.org/abs/2205.03414}{{\tt arXiv:2205.03414}}.

\bibitem{Low:1958sn}
F.~E. Low, {\it {Bremsstrahlung of very low-energy quanta in elementary
  particle collisions}},  {\em Phys. Rev.} {\bf 110} (1958) 974--977.

\bibitem{Burnett:1967km}
T.~H. Burnett and N.~M. Kroll, {\it {Extension of the low soft photon
  theorem}},  {\em Phys. Rev. Lett.} {\bf 20} (1968) 86.

\bibitem{DelDuca:1990gz}
V.~Del~Duca, {\it {High-energy Bremsstrahlung Theorems for Soft Photons}},
  {\em Nucl. Phys.} {\bf B345} (1990) 369--388.

\bibitem{Moult:2016fqy}
I.~Moult, L.~Rothen, I.~W. Stewart, F.~J. Tackmann, and H.~X. Zhu, {\it
  {Subleading Power Corrections for N-Jettiness Subtractions}},
  \href{http://arxiv.org/abs/1612.00450}{{\tt arXiv:1612.00450}}.

\bibitem{Boughezal:2016zws}
R.~Boughezal, X.~Liu, and F.~Petriello, {\it {Power Corrections in the
  N-jettiness Subtraction Scheme}},
  \href{http://arxiv.org/abs/1612.02911}{{\tt arXiv:1612.02911}}.

\bibitem{Moult:2017jsg}
I.~Moult, L.~Rothen, I.~W. Stewart, F.~J. Tackmann, and H.~X. Zhu, {\it
  {N-Jettiness Subtractions for $gg\to H$ at Subleading Power}},
  \href{http://arxiv.org/abs/1710.03227}{{\tt arXiv:1710.03227}}.

\bibitem{Boughezal:2018mvf}
R.~Boughezal, A.~Isgr{\`o}, and F.~Petriello, {\it {Next-to-leading-logarithmic
  power corrections for $N$-jettiness subtraction in color-singlet
  production}},  \href{http://arxiv.org/abs/1802.00456}{{\tt
  arXiv:1802.00456}}.

\bibitem{Moult:2018jjd}
I.~Moult, I.~W. Stewart, G.~Vita, and H.~X. Zhu, {\it {First Subleading Power
  Resummation for Event Shapes}},  \href{http://arxiv.org/abs/1804.04665}{{\tt
  arXiv:1804.04665}}.

\bibitem{Jaffe:1982pm}
R.~L. Jaffe and M.~Soldate, {\it {Twist Four in Electroproduction: Canonical
  Operators and Coefficient Functions}},  {\em Phys. Rev.} {\bf D26} (1982)
  49--68.

\bibitem{CMS:2013kfa}
{\bf CMS} Collaboration, {\it {Performance of quark/gluon discrimination in 8
  TeV pp data}}, .

\bibitem{ATLAS:2016wzt}
{\bf ATLAS} Collaboration, {\it {Discrimination of Light Quark and Gluon Jets
  in $pp$ collisions at $\sqrt{s} = 8$ TeV with the ATLAS Detector}}, .

\bibitem{Larkoski:2014wba}
A.~J. Larkoski, S.~Marzani, G.~Soyez, and J.~Thaler, {\it {Soft Drop}},  {\em
  JHEP} {\bf 05} (2014) 146, [\href{http://arxiv.org/abs/1402.2657}{{\tt
  arXiv:1402.2657}}].

\bibitem{Medves:2022ccw}
R.~Medves, A.~Soto-Ontoso, and G.~Soyez, {\it {Lund and Cambridge
  multiplicities for precision physics}},
  \href{http://arxiv.org/abs/2205.02861}{{\tt arXiv:2205.02861}}.

\bibitem{dire}
S.~H\"oche and S.~Prestel, {\it {The midpoint between dipole and parton
  showers}},  {\em Eur. Phys. J. C} {\bf 75} (2015), no.~9 461,
  [\href{http://arxiv.org/abs/1506.05057}{{\tt arXiv:1506.05057}}].

\bibitem{pythia82}
T.~Sj\"ostrand, S.~Ask, J.~R. Christiansen, R.~Corke, N.~Desai, P.~Ilten,
  S.~Mrenna, S.~Prestel, C.~O. Rasmussen, and P.~Z. Skands, {\it {An
  introduction to PYTHIA 8.2}},  {\em Comput. Phys. Commun.} {\bf 191} (2015)
  159--177, [\href{http://arxiv.org/abs/1410.3012}{{\tt arXiv:1410.3012}}].

\bibitem{Cacciari:2008gp}
M.~Cacciari, G.~P. Salam, and G.~Soyez, {\it {The Anti-k(t) jet clustering
  algorithm}},  {\em JHEP} {\bf 04} (2008) 063,
  [\href{http://arxiv.org/abs/0802.1189}{{\tt arXiv:0802.1189}}].

\bibitem{NIPS2017_f22e4747}
M.~Zaheer, S.~Kottur, S.~Ravanbakhsh, B.~Poczos, R.~R. Salakhutdinov, and A.~J.
  Smola, {\it Deep sets},  in {\em Advances in Neural Information Processing
  Systems} (I.~Guyon, U.~V. Luxburg, S.~Bengio, H.~Wallach, R.~Fergus,
  S.~Vishwanathan, and R.~Garnett, eds.), vol.~30, Curran Associates, Inc.,
  2017.

\bibitem{Komiske:2018cqr}
P.~T. Komiske, E.~M. Metodiev, and J.~Thaler, {\it {Energy Flow Networks: Deep
  Sets for Particle Jets}},  {\em JHEP} {\bf 01} (2019) 121,
  [\href{http://arxiv.org/abs/1810.05165}{{\tt arXiv:1810.05165}}].

\bibitem{Komiske:2017aww}
P.~T. Komiske, E.~M. Metodiev, and J.~Thaler, {\it {Energy flow polynomials: A
  complete linear basis for jet substructure}},
  \href{http://arxiv.org/abs/1712.07124}{{\tt arXiv:1712.07124}}.

\bibitem{tensorflow}
M.~Abadi, P.~Barham, J.~Chen, Z.~Chen, A.~Davis, J.~Dean, M.~Devin,
  S.~Ghemawat, G.~Irving, M.~Isard, et~al., {\it Tensorflow: A system for
  large-scale machine learning.},  in {\em OSDI}, vol.~16, pp.~265--283, 2016.

\bibitem{keras}
F.~Chollet, ``Keras.'' \url{https://github.com/fchollet/keras}, 2017.

\bibitem{adam}
D.~Kingma and J.~Ba, {\it Adam: A method for stochastic optimization},
  \href{http://arxiv.org/abs/1412.6980}{{\tt arXiv:1412.6980}}.

\bibitem{Larkoski:2014pca}
A.~J. Larkoski, J.~Thaler, and W.~J. Waalewijn, {\it {Gaining (Mutual)
  Information about Quark/Gluon Discrimination}},  {\em JHEP} {\bf 11} (2014)
  129, [\href{http://arxiv.org/abs/1408.3122}{{\tt arXiv:1408.3122}}].

\end{thebibliography}\endgroup

\end{document}